\documentclass[conference]{IEEEtran}
\usepackage{cite}
\usepackage{amsmath,amssymb,amsfonts}
\usepackage{algorithmic}
\usepackage{graphicx}
\usepackage{textcomp}
\usepackage{xcolor}

\usepackage{hyperref}
\usepackage{url}
\usepackage{subcaption}
\usepackage{makecell}
\usepackage{multirow}

\AtBeginDocument{%
  \providecommand\BibTeX{{%
    Bib\TeX}}}
\def\BibTeX{{\rm B\kern-.05em{\sc i\kern-.025em b}\kern-.08em
    T\kern-.1667em\lower.7ex\hbox{E}\kern-.125emX}}

\begin{document}

\title{Scalable Benchmarking Framework for Dynamic Quantum Circuits}

\author{\IEEEauthorblockN{Sumeet Shirgure}
\IEEEauthorblockA{
University of Central Florida\\
Orlando, USA \\
sumeetparameshwar.shirgure@ucf.edu}\\
\and
\IEEEauthorblockN{Efekan K\"okc\"u}
\IEEEauthorblockA{University of Central Florida\\
Orlando, USA \\
efekan.kokcu@ucf.edu}
\and
\IEEEauthorblockN{Anupam Mitra}
\IEEEauthorblockA{Lawrence Berkeley National Laboratory\\
Berkeley, USA \\
AnupamMitra@lbl.gov}
\and
\IEEEauthorblockN{Wibe Albert de Jong}
\IEEEauthorblockA{Lawrence Berkeley National Laboratory\\
Berkeley, USA \\
wadejong@lbl.gov}
\and
\IEEEauthorblockN{Costin Iancu}
\IEEEauthorblockA{Lawrence Berkeley National Laboratory\\
Berkeley, USA \\
CCIancu@lbl.gov}
\and
\IEEEauthorblockN{Siyuan Niu}
\IEEEauthorblockA{
\textit{University of Central Florida}\\
Orlando, USA \\
siyuan.niu@ucf.edu}
}

\maketitle

\begin{abstract}
  \label{sec:abstract}
  Dynamic quantum circuits with mid-circuit measurements (MCMs) and feed-forward operations play a crucial role in various applications, such as quantum error correction and quantum algorithms. With advancements in quantum hardware enabling the implementation of MCM and feed-forward loops, the use of dynamic circuits has become increasingly prevalent. There is a significant need for a benchmarking framework specially designed for dynamic circuits to capture their unique properties, as current benchmarking tools are designed primarily for unitary circuits and cannot be trivially extended to dynamic circuits. We propose \textit{dynamarq}, a scalable and hardware-agnostic benchmarking framework for dynamic circuits. We collect a suite of dynamic circuit benchmarks spanning various applications and propose a broad set of circuit features to characterize the structure of these dynamic circuits. We run them on two IBM quantum processors and the Quantinuum Helios-1E emulator, and propose scalable, application-dependent fidelity scores for each benchmark based on hardware execution results. We perform statistical modeling to identify correlations between circuit features and fidelity scores, and demonstrate highly accurate fidelity prediction using our model. Our model parameters are also transferable across hardware backends and calibration cycles. Our framework facilitates the understanding of dynamic circuit structures and provides insights for designing and optimizing dynamic circuits to achieve high execution fidelity on quantum hardware.
\end{abstract}


\section{Introduction}
  \label{sec:introduction}

Dynamic quantum circuits incorporate mid-circuit measurements (MCMs) and feed-forward operations. They are a key component in quantum error correction (QEC) and quantum teleportation. Beyond these foundational applications, dynamic circuits offer significant advantages in reducing resource overhead. For example, they enable the preparation of quantum states such as GHZ and matrix product states, and the implementation of specific unitaries such as long-range CNOTs, with constant or significantly reduced circuit depths \cite{smith2023deterministic, smith2024constant, baumer2025measurement, baumer2024efficient, farrell2025digital, zi2025constant, alam2024learning, buhrman2024state,niu2024acdcautomatedcompilationdynamic}, and reduce the number of two-qubit gates in algorithms like Quantum Fourier Transform (QFT) \cite{baumer2024quantum,baumer2025approximate}. By leveraging MCM and reset, dynamic circuits can also minimize overall circuit width~\cite{niu2024effective, hua2023caqr, decross2023qubit, fang2025dynamic}. Recent advances demonstrate that certain dynamic-circuit based quantum algorithms provide substantial advantages over unitary circuits~\cite{cao2026measurement, lemelin2025mid,foss2023experimental}.
On the hardware side, MCM and feed-forward operations have been successfully realized across various quantum platforms~\cite{koh2023measurement, xu2023qubic,pino2021demonstration,lis2023midcircuit}, bridging the gap between theoretical proposals and practical implementation. Together, these algorithmic and hardware advancements indicate that quantum applications leveraging dynamic circuits will become increasingly prevalent. Therefore, benchmarking the capability of real quantum hardware to execute dynamic circuits has become timely and necessary.

However, no dedicated benchmarking framework exists for dynamic circuits, as current methods are primarily designed for conventional unitary (static) circuits where measurements are performed only at the end of the circuit~\cite{patel2025platform,lubinski2023application, supermarq, li2023qasmbench, quetschlich2023mqt, cosentino2026metriq}. Naively extending existing benchmarking frameworks for unitary circuits to dynamic circuits is insufficient for four key reasons: (1) New benchmark datasets must be collected based on existing dynamic circuit applications, as no dedicated benchmarking suites currently exist. (2) Dynamic circuit structures are non-deterministic, changing based on measurement outcomes, which existing benchmarking frameworks fail to capture. (3) Information before MCMs propagates to the gates after feed-forward operations, and this dependency is overlooked by existing frameworks. (4) Dynamic circuits introduce distinct hardware noise sources from MCM and feed-forward operations, including MCM errors and decoherence errors introduced by the extended duration of MCM and feed-forward operations.
Therefore, a new benchmarking framework dedicated to dynamic circuits is required.

To address the gap, we propose a hardware-agnostic dynamic circuit benchmarking framework called \textit{dynamarq}. While a benchmarking framework can be developed at the component, system, or application level, in this work we focus on the application level, aiming to evaluate quantum hardware capability in executing dynamic circuit applications. \textit{dynamarq} includes a dataset of dynamic circuit benchmarks spanning quantum state preparation, quantum algorithms, quantum gates, and QEC. We introduce a set of features to analyze circuit structures with a particular focus on capturing the dynamic aspects of the circuits. The benchmarks are executed on two IBM quantum hardware with different architectures and the Quantinuum Helios Emulator (known to have a very accurate noise model compared to real quantum hardware \cite{montanez2025evaluating}). Based on the hardware execution results, we propose an application-dependent fidelity score for each circuit type that is scalable to an arbitrary number of qubits. We then perform statistical modeling to identify the features most strongly correlated with hardware fidelity as well as predicting the circuit fidelity scores based on the circuit features. Our model achieves up to 53.4\% and 2.8$\times$ higher predictive capability than the state of the art on IBM quantum hardware and the Quantinuum Helios Emulator.  Moreover, since MCM and feed-forward operations introduce additional decoherence errors, dynamical decoupling (DD)~\cite{souza2011robust,niu2022effects,lidar2014review} is commonly used to suppress such noise.
We demonstrate that our fidelity prediction framework remains robustly effective when DD error mitigation techniques are applied. Finally, we show that the statistical modeling parameters are transferable across quantum hardware within the same qubit modality and across calibration cycles, for correlation analysis and fidelity prediction.

The main contributions of the paper are as follows :

\begin{itemize}
    \item To the best of our knowledge, we propose the first dynamic circuit benchmarking framework, \textit{dynamarq}, comprising a suite of benchmarks focused on various dynamic circuits applications.

    \item We introduce a broad set of circuit features tailored to dynamic circuits, particularly focusing on the impact of MCM, feed-forward classical control, and the dynamic structure of circuits that depends on the MCM outcomes.

    \item We execute our benchmarks on two IBM quantum hardware systems with different architectures and the Quantinuum Helios Emulator, with and without error mitigation settings, and propose application-specific, scalable fidelity scores for quantum circuits of arbitrary size based on the hardware execution results.

    \item We perform statistical modeling to characterize the correlations between circuit features and fidelity scores, and demonstrate high-accuracy circuit fidelity prediction from circuit features, across different quantum hardware systems and the emulator, with and without error mitigation.
    
    \item We show that the model parameters are transferable across quantum hardware systems within the same qubit modality and remain robust across different calibration cycles.
\end{itemize}

We have open-sourced \textit{dynamarq} \cite{dynamarq} as a PyPI Python package that provides hardware-agnostic circuit generators using Qiskit \cite{javadi2024quantum} for IBM quantum computers and Guppy \cite{koch2025guppy} for Quantinuum systems.

\textit{dynamarq} is designed to be easily extensible and open source, allowing users to integrate their own dynamic circuit applications and perform fidelity and circuit feature analysis using our statistical models.
It provides circuit generators in both Qiskit \cite{javadi2024quantum} and Guppy \cite{koch2025guppy}.
The resulting analysis reveals which circuit features most significantly impact hardware execution fidelity, providing insights for designing and optimizing dynamic circuit structures to achieve higher fidelity on real quantum hardware. 
\section{Background}
  \label{sec:background}

\subsection{Dynamic Circuits}
Unlike unitary circuits, where measurements are only performed at the
end after all the gates have been completed, dynamic circuits allow for
the measurement of qubits at intermediate steps.
Based on the results of these measurements, additional operations can
be dynamically applied to the unmeasured qubits.
An example is shown in Figure~\ref{fig:motivation}: we apply MCM to $q_1$, and the measurement result decides the circuit structure, where $X$ is applied to $q_2$ only if the measurement result is 1.  
We refer to the portion of the circuit before the MCM as the \textit{base} circuit and the portion after the MCM as the \textit{branch} circuit.

While dynamic circuits offer various advantages over unitary circuits such as resource reduction and computational advantages, a key challenge is the substantial errors introduced by MCMs and feed-forward operations. For example, on IBM Kingston, the median (final) measurement errors are five times higher than two-qubit gate errors, with measurement durations that are 34 times longer. IBM recently introduced a new MCM feature achieving a $65\%$ reduction in measurement duration compared to the final measurement, though at an error rate ten times that of two-qubit gates. Beyond measurement latency, feed-forward operations incur additional classical communication latency of approximately 600 $ns$ on IBM hardware. MCMs can also introduce errors on unmeasured spectator qubits through measurement-induced dephasing~\cite{hashim2025efficient}. Thus, while dynamic circuits currently face a trade-off between their advantages and increased noise, their benefits will become increasingly pronounced as hardware fidelity improves.

\subsection{Prior Work}
Quantum computers based on various qubit modalities have made remarkable progress over the past decade, making benchmarking tools essential for tracking advances across different metrics and abstraction levels. Current quantum benchmarking protocols are broadly categorized into three levels: (1) Component-level hardware characterization, such as Randomized Benchmarking~\cite{PhysRevLett.102.090502}, Cycle Benchmarking~\cite{erhard2019characterizing}, and Cross-Entropy Benchmarking~\cite{boixo2018characterizing}; (2) System-level performance evaluation, such as Quantum Volume~\cite{PhysRevA.100.032328} and Circuit Layer Operations Per Second (CLOPS)~\cite{wack2021qualityspeedscalekey}; and (3) Application-level performance analysis, such as SupermarQ~\cite{supermarq}, QUARK~\cite{finvzgar2022quark}, BACQ~\cite{barbaresco2025bacq}, and QED-C Application-Oriented Benchmarks~\cite{lubinski2023application}. Several software tools and frameworks support multiple levels of these benchmarking protocols, such as pyGSTi~\cite{Nielsen2020-rd} and Metriq~\cite{cosentino2026metriq}. In addition, a number of benchmarking suites have been proposed, such as Benchpress~\cite{nation2025benchmarking}, MQT Bench~\cite{quetschlich2023mqtbench}, and QASMBench~\cite{li2023qasmbench}. Beyond benchmarking, a complementary line of work focuses on predicting circuit fidelity directly from hardware calibration data or specific noise models and circuit structure, leveraging machine learning models to estimate performance on specific quantum devices~\cite{vadali2024quantum, mao2025q, saravanan2022machine}. However, all aforementioned efforts, spanning both benchmarking protocols and fidelity prediction methods, are designed primarily for unitary circuits.

For dynamic circuits, benchmarking efforts remain limited: protocols have been proposed for quantifying errors on MCM qubits and spectator qubits~\cite{Govia_2023}, and some works propose integrating dynamic circuits into Randomized Benchmarking~\cite{PhysRevA.111.012611, hothem2025measuring}. These efforts, however, focus exclusively on the component level, and no application-level benchmarking protocols for dynamic circuits currently exist.
\section{Motivation}

\begin{figure}
    \centering
    \includegraphics[width = 0.25\textwidth]{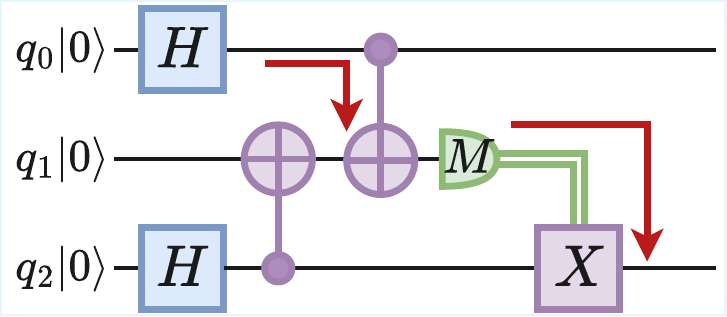}
    \caption{A dynamic circuit that prepares 2-qubit GHZ state.
    Note that the information propagates (red arrow) from $q_1$ to $q_2$
    through MCM and feed-forward operations.}
    \label{fig:motivation}
\end{figure}

Figure~\ref{fig:motivation} shows a two-qubit GHZ state (Bell state), with $q_0, q_2$ as data (system) qubits, and $q_1$ as ancilla qubit that has an MCM. The existing tools will analyze the circuit structure in a "unitary" way: for example, using Qiskit~\cite{javadi2024quantum} to compute the circuit depth yields a value of 5, accounting for the base circuit depth, the MCM, and the $X$ gate on the branch circuit. Moreover, computing the number of entangling gates yields 2, corresponding to the two CNOT gates.

However, we argue that the structure of dynamic circuits must be analyzed dynamically, since the circuit structure changes according to the measurement result. Take circuit depth as an example: the actual depth depends on the measurement outcome, where the maximum depth is 5 when the measurement result is 1, and the minimum depth is 4 when it is 0. Assuming equal measurement probability, the \textbf{expected} depth should be 4.5. This is only a toy example; as branch circuits become more complex, the difference in depth across branches will be more significant. However, obtaining exact measurement probabilities requires hardware execution or classical simulation, which we aim to avoid. Therefore, we propose efficient approximation methods, detailed in Section~\ref{sec:prob}. The concept of expected values can also be extended to other circuit features, such as gate count. For example, whether the $X$ gate on $q_2$ is counted depends on the measurement result, so we should use the expected number of gates instead. Furthermore, when calculating entangling gates, we must account for entanglement both before and after MCM. Before the MCM, the circuit contains 2 CNOTs. After the MCM, even though the $X$ gate is considered a classically-controlled single-qubit gate, it effectively depends on the MCM result of $q_1$, since information from the base circuit propagates into the branch circuits as shown by the red arrow. We refer to this propagated information as \textbf{``feed-forward correlation"},
which must be captured and included when calculating the overall entanglement. Finally, since feed-forward operations can have longer durations than standard quantum gates depending on the hardware, we must carefully decide whether to include them in circuit depth or operation count. These considerations motivate us to rethink how circuit features are defined and measured for dynamic circuits.
\section{Overview of Proposed Benchmarking Framework}
Figure~\ref{fig:overview} shows the overview of our proposed application-level dynamic circuit benchmarking framework \textit{dynamarq}. 
The essential components of our framework are summarized below:

\begin{figure}[h]
    \centering
    \includegraphics[width=0.8\linewidth]{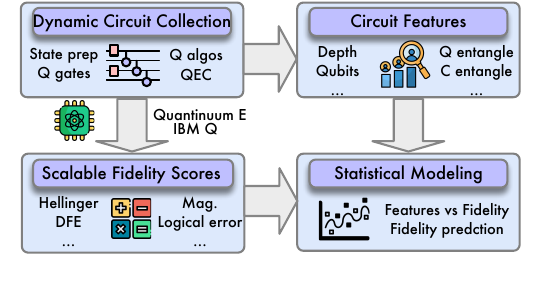}
    \caption{Overview of the dynamic circuit benchmarking framework.}
    \label{fig:overview}
\end{figure}

1) \textbf{Dynamic circuit collection}.
A collection of dynamic circuit benchmarks spanning state preparation, quantum gate implementations, quantum algorithms, and QEC. These benchmarks are provided as parameterized functions, allowing users to generate circuits of arbitrary size.

2) \textbf{Circuit features}. We introduce a set of circuit features to characterize dynamic circuits, categorized into unnormalized features (e.g., gate count and circuit depth) and normalized features (e.g., quantum entanglement and ``feed-forward correlation"),
capturing information across three key components: the base circuit, the MCM and feed-forward control, and the probabilistic branch circuits.

3) \textbf{Scalable application-dependent fidelity scores}. We execute the benchmarks on quantum hardware or emulators and develop application-dependent fidelity scores by comparing execution results against ideal outcomes. The fidelity score calculation is scalable to arbitrarily large quantum circuits.

4) \textbf{Statistical modeling}. We perform statistical modeling to analyze the correlations between circuit features and fidelity scores using a linear regression model. This enables us to identify which circuit features most significantly impact fidelity, as well as to predict the fidelity score of a given circuit from its features.
\section{Dynamic Circuit Features}
\label{sec:features}

To propose features dedicated to dynamic circuit properties, we address three key challenges.   First, we define \textit{expected} feature values that account for measurement outcome probabilities, where the probabilities are efficiently approximated without classical simulation (Section~\ref{sec:prob}).

Second, we introduce \textit{feed-forward correlation} to capture dependencies between branch circuits and MCM qubits. Third, we incorporate MCM and feed-forward operations into key feature counts; however, since their impact is highly hardware-dependent, we introduce multiple feature design variants to ensure flexibility as both quantum and classical hardware evolve.
The overall features are categorized as unnormalized (Section~\ref{sec:unnormalized-metrics}) and normalized (Section~\ref{sec:normalized-metrics}).

\subsection{The Choice of Branch Probability}\label{sec:prob}
\begin{figure}
    \centering
    \includegraphics[width=0.35\textwidth]{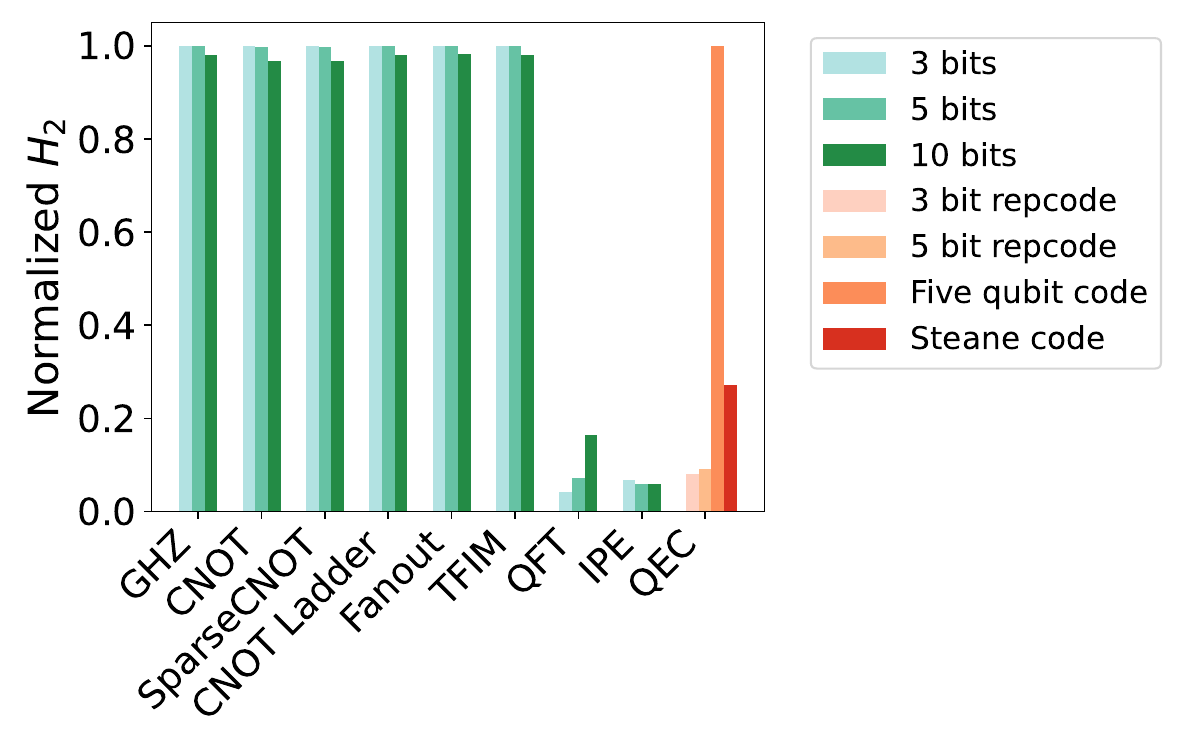}
    \caption{
    The normalized R\'{e}nyi-2 entropy $H_2$ of the probability distributions of mid-circuit measurement outcomes on IBM Pittsburgh. A high score means the probability distribution is nearly uniform across all measurement outcomes. The entropy is normalized by the number of classical bits as $H_2 / n_a$.
    Details of these benchmarks are in Section \ref{sec:benchmark_suite}.
   }
    \label{fig:branch_probabilities}
\end{figure}
The branch probability is required to calculate the expected feature values introduced in the following sections. However, estimating the branch probabilities for the benchmarks introduced in Section~\ref{sec:benchmark_suite} is non-trivial, as we aim to avoid simulating the dynamic circuit to determine their exact values. For most benchmarks, based on the observation in~\cite{niu2024acdcautomatedcompilationdynamic}, the branch probability is uniformly distributed, and we therefore set $p_i = \frac{1}{2^{n_a}}$, where $n_a$ is number of classical bits (obtained from MCM results) on which the feed forward operation is conditioned. For benchmarks such as Quantum Fourier Transform (QFT) and Iterative Phase Estimation (IPE), the branch probability depends on the input string. Since the input string is randomly selected, averaging over all possible input strings also yields a uniform distribution. For syndrome measurement circuits in QEC codes, however, the branch probability corresponds to the correction unitaries, which depend on the circuit-level noise of the underlying hardware, and therefore requires a dedicated estimation method.  To verify our estimation, Figure~\ref{fig:branch_probabilities} plots the normalized Rényi-2 entropy $H_2/n_a$ of MCM outcome distributions across benchmarks executed on IBM Pittsburgh,
where $H_2 = -\log_2(\sum_i p_i^2)$.
An entropy close to one means the observed branch statistics
are uniformly distributed, whereas lower values indicate a skew toward some preferred branches.
The results confirm our estimation: (1) most benchmarks have a normalized entropy of one, consistent with a uniform branch probability distribution; (2) QFT and IPE exhibit lower entropy as they are evaluated on a single input bit string; and (3) QEC codes have low entropy because their output distribution is peaked at the all zero (no-correction) branch (except the 5-qubit code). In summary, we assign equal branch probabilities to all benchmarks except QEC codes, for which we elaborate on a dedicated estimation method and detail the 5-qubit code exception in Section~\ref{sec:ecc}.

\subsection{Unnormalized Features}
\label{sec:unnormalized-metrics}

\subsubsection{Total Circuit Depth ($D_{\text{total}})$}

\label{sec:depth}

The depth of a circuit is defined as the minimum number of sequential gate layers required to execute a quantum circuit, assuming parallel gate execution is possible.
For a dynamic circuit, the depth $D_{\text{total}}$ is comprised of four components, each of which is defined as follows :
\begin{itemize}
    \item $D_{\text{base}}$ : the depth associated with the base circuit, which remains the same across MCM results.
    \item $D_{\text{classical}}$ : the depth associated with feed-forward operations. Since they introduce a latency that varies across hardware modalities,
    whether to include these operations is left as a user-defined option, controlled by a flag. Note that feed-forward operations may also involve additional classical logic such as XORs and switch-like operations; however, since these are generally much faster than quantum operations and are hardware-dependent, we exclude them from our analysis.
    \item $D_{\text{MCM}}$ : the depth associated with MCM operations. We increment the depth by one for every MCM operation on the critical path.
    \item $D_{\text{branch}} = \sum_{i} p_i \cdot D_i$ : the expected branch depth weighted by execution probabilities, where $p_i$ is the probability of executing the $i$-th branch circuit and $D_i$ is its corresponding depth. The choice of $p_i$ is non-trivial, as we aim to avoid simulating the dynamic circuit to determine its exact value. The estimation of $p_i$ is elaborated in Section~\ref{sec:prob}.
\end{itemize}

\subsubsection{Total Number of Operations ($O_{\text{total}}$)}
When calculating the total number of operations, we consider contributions from the base circuit, branch circuits, and classical processing. Circuit operations are categorized as either unitary or non-unitary, where non-unitary operations refer to measurements and resets. We introduce the following notation:
\begin{itemize}
\item $U_{\text{base}}$ and $O_{\text{base}}$: the number of unitary operations and total operations (unitary and non-unitary) in the base circuit, respectively.
\item $U_{\text{branch}}$ and $O_{\text{branch}}$: the expected unitary and total operation counts over branch circuits, respectively:
$        U_{\text{branch}} = \sum_{i} p_i \cdot U_i, \quad O_{\text{branch}} = \sum_{i} p_i \cdot O_i,
   $
where $p_i$ is the probability of executing the $i$-th branch circuit, and $U_i$ and $O_i$ are the unitary and total operation counts of the $i$-th branch circuit, respectively.
\item $O_{\text{classical}}$: the number of feed-forward operations, counted as 1 regardless of the classical logic involved.
\end{itemize}
Based on these, we define three variants of the total number of operations $O_{\text{total}}$:

\begin{equation}
O_{\text{total}} = 
\begin{cases} 
U_{\text{base}} + U_{\text{branch}} & \text{(unitary only)} \\
O_{\text{base}} + O_{\text{branch}} & \text{(all quantum operations)} \\
O_{\text{base}} + O_{\text{branch}} + O_{\text{classical}} & \text{(+ classical processing)}
\end{cases}
\end{equation}

\subsubsection{Total Number of Qubits ($n_{\text{total}}$)} The total number of qubits in the circuit, including ancilla qubits used for MCMs. For example, implementing a long-range CNOT requires intermediate ancilla qubits in addition to the control and target qubits, all of which are counted in $n_{\text{total}}$. We also define a variant $n_s$ that counts only the information-carrying (system) qubits.

\subsection{Normalized Features}
\label{sec:normalized-metrics}
We adopt some feature concepts from SupermarQ~\cite{supermarq} and propose dedicated adaptations for dynamic circuits, while also introducing new features that capture their unique properties. All feature values are normalized to $[0, 1]$.

\subsubsection{Quantum Entanglement}
The quantum entanglement is defined as the ratio of two-qubit gates in the base circuit $g_{2q}^{\text{base}}$ to the total number of operations $O_{\text{total}}$:

\begin{equation}
    E_Q = \frac{g_{2q}^{\text{base}}}{O_\text{total}}
\end{equation}
Note that only two-qubit gates from the base circuit are counted in the numerator, since branch circuits are executed only for certain measurement outcomes. Including branch circuit gates would not accurately reflect the unconditional entanglement generated by the circuit, as $E_Q$ is intended to measure the proportion of entangling operations that are always present, regardless of dynamic branches.

\subsubsection{Quantum Entanglement + Feed-forward Correlation}
Feed-forward Correlation refers to the correlation created between the MCM qubit and the gates in the branch circuits, conditioned on the measurement outcome.

For example, consider the dynamic circuit shown in Figure~\ref{fig:motivation}, where $q_1$ is measured: if the outcome is 1, an $X$ gate is applied to $q_2$; otherwise, no gate is applied. This process creates correlations between $q_2$ and $q_1$ with probability $p_1$, where $p_1$ is the probability that the measurement outcome of $q_1$ is 1. Similarly, if a CNOT gate appears in the branch circuit, it is effectively a Toffoli gate, as it is conditioned on both the MCM outcome and the control qubit in the branch subcircuit.
In this paper, for simplicity, we count such a gate as one unit of
feed-forward correlation,
such that all gates in the branch circuits are considered
correlated with the MCM qubit.
A more accurate representation of such gates is left for future work.
To summarize, feed-forward correlation is defined as the ratio of the expected number of gates in the branch circuits to the total number of operations in the entire circuit:
\begin{equation}
E_C = \frac{\sum_{i} p_i \cdot g_{i}^{\text{branch}}}{O_\text{total}}
\end{equation}
\noindent where $p_i$ is the probability of the $i$-th MCM outcome and $g_{i}^{\text{branch}}$ is the number of gates in the corresponding branch circuit. The total entanglement, combining both quantum and feed-forward correlation contributions, is then defined as:
\begin{equation}
E_{QC} = E_Q + E_C.
\end{equation}

\subsubsection{System-Qubit-Ratio}
In dynamic circuits, qubits are categorized into two types: (1) system (data) qubits, which carry the useful information during the computation; and (2) ancilla qubits, used for MCMs and feed-forward operations applied to the system qubits. For example, in quantum state preparation, system qubits encode the desired state while ancilla qubits serve as intermediate resources that facilitate circuit implementation via MCM and feed-forward operations; in QEC, system qubits encode the logical qubit while ancilla qubits are used to detect errors.

The system-qubit-ratio provides a measure of the proportion
of useful (information-carrying) qubits to the total number of qubits:
    \begin{equation}
       SQ  = \frac{n_s}{n}
    \end{equation}
where $n_s$ is the number of system qubits and $n$ is the total
number of qubits.

\subsubsection{Two-qubit Critical Depth}

The critical path is the longest path in a circuit that determines its depth and execution time. Since two-qubit gates typically dominate execution time, the two-qubit critical depth $\eta_{2q}$ is defined as the ratio of two-qubit gates on the critical path to the total number of two-qubit gates. If this ratio is close to one, it indicates that two-qubit gates
are executed in a highly serial manner. For dynamic circuits, the contributions from two-qubit gates in both base and  branch circuits must be considered.
The two-qubit critical depth is calculated as:

\begin{equation}
  \eta_{2q} = \frac{N_{2q,\text{crit}}^{\text{base}} +
   \sum_{i} p_i N_{2q,\text{crit}}^{(i)}}{N_{2q,\text{total}}^{\text{base}}
   + \sum_{i} p_i N_{2q,\text{total}}^{(i)}}
\end{equation}
where $N_{2q,\text{crit}}^{\text{base}}$ and $N_{2q,\text{total}}^{\text{base}}$ are the number of two-qubit gates on the critical path and the total number of two-qubit gates in the base circuit, respectively; $N_{2q,\text{crit}}^{(i)}$ and $N_{2q,\text{total}}^{(i)}$ are the corresponding quantities for the $i$-th branch circuit; and $p_i$ is the probability of taking branch $i$.

\subsubsection{Dynamic Depth Ratio} 
The dynamic depth ratio quantifies the contribution of key dynamic features (MCMs and feed-forward operations), to the overall circuit execution, representing the fraction of circuit depth occupied by these dynamic operations. Since feed-forward operations are highly hardware-dependent, the dynamic depth ratio can be defined in two ways configured by a flag: if the flag is set to \texttt{True}, the dynamic depth includes both MCM and feed-forward layers; otherwise, it only accounts for MCM layers. Formally, the dynamic depth ratio is defined as:
\begin{equation}
    D_\text{dynamic} =
\begin{cases}
    \frac{l_{\text{mcm}} + l_{\text{ff}}}{l_{\text{total}}},
    & \text{if flag is True} \\
    \frac{l_{\text{mcm}}}{l_{\text{total}}}, & \text{if flag is False}
\end{cases}
\end{equation}
where $l_{\text{mcm}}$ is the number of mid-circuit measurement layers,
$l_{\text{ff}}$ is the number of feed-forward layers,
and $l_{\text{total}}$ is the total number of layers.

\subsubsection{Program Communication}
The program communication metric captures the qubit interactions in the circuit.
A straightforward approach is to represent the circuit as a graph,
where each node corresponds to a qubit.
An edge is created between two nodes if there is an entangling gate between
the corresponding qubits, and the degree of each node indicates how
many other qubits it interacts with.
In dynamic circuits, feed-forward correlation
can induce interactions between qubit pairs, resulting in edges that may only exist with certain probabilities.
To capture this, we construct a weighted connectivity matrix $A$, where $A_{ij}$ is the probability that we apply two-qubit gate between qubits $i$ and $j$.
Note that if there is a single-qubit gate on $i$ conditioned
on the MCM outcome of $j$ we consider that that branch contributes
to $A_{ij}$ at a rate given by the corresponding branch probability.
We make a simplifying assumption that different branch instructions have
independent probability distributions, which is the case for all of the
benchmarks in our suite.
The degree of node $i$ in $A$ is given by $d_i = \sum_{j \neq i} A_{ij}$.
The normalized program communication for the entire circuit is defined as:
\begin{equation}
    C = \frac{1}{n(n-1)} \sum_{i} d_i
\end{equation}
where $n$ is the total number of qubits in the circuit,
and $d_i$ is the degree of qubit $i$.
We refer to $A$ computed just using the base circuit as \textit{quantum communication},
and the $A$ which includes branch circuits as \textit{quantum + classical communication}.

\subsubsection{Liveness}
During the execution of a quantum circuit, qubits are considered
"live" when they are actively involved in gate operations, and "idle"
when they are waiting for other qubits to complete their operations.
The liveness of a quantum circuit reflects the proportion of time
that qubits are active during execution.
Low liveness indicates longer idle periods, during which qubits are
more susceptible to decoherence errors.
To quantify the liveness of a quantum circuit, we represent each
qubit's total execution time by the circuit depth, and define a qubit's
live time as the number of time steps in which it participates in a gate.
The overall liveness of the circuit is then given by the ratio of the
total live time across all qubits to the total execution time.
The total live time for a dynamic circuit is calculated as:
\begin{align}
    T_{\text{live}} =\ & g_{1q} ^{\text{base}} +
        2 \times g_{2q}^{\text{base}} +
        g_{\text{reset}} ^{\text{base}} \notag \\
    & + \sum_{i} p_i \cdot (g_{i, 1q} ^{\text{branch}}
        + 2 \times g_{i, 2q}^{\text{branch}}
        + g_{i, \text{reset}} ^{\text{branch}})
\end{align}
where $g_{1q}$, $g_{2q}$ and $g_{\text{reset}}$ represents the number of
single-qubit gates, two-qubit gates, and reset operations.
The factor 2 before $g_{2q}$ accounts for the fact that a two-qubit gate
takes one time step for two qubits involved. 
To calculate the total execution time, we first preprocess the circuit
by removing the final measurements.
The execution time for each qubit is categorized as follows:
(1) For qubits that undergo MCMs and are not subsequently reset and reused
(denoted as $n_1)$, we use the circuit depth preceding the MCM to represent
their execution time.
(2) For qubits that undergo MCMs and are reset and reused, as well as
other qubits without MCM (denoted as $n_2$), we use the total circuit
depth $D_{\text{total}}$ as defined in
Section \ref{sec:unnormalized-metrics}.
The total execution time is calculated as:
\begin{equation}
    T_\text{execution}  = \sum_{i=0}^{n_1} D_{i, \text{pre}}
        + n_2 \cdot D_\text{total}
\end{equation}
where $D_{i, \text{pre}}$ is the circuit depth preceding the
final measurement for qubit $q_i$.
The liveness $L$ of the circuit is:

\begin{equation}
    L = \frac{T_\text{live}}{T_\text{execution}}
\end{equation}

\subsubsection{Parallelism}
The parallelism feature indicates the extent to which gate operations
are executed simultaneously on different qubits.
In general, quantum circuits are compiled to maximize parallel
execution of gates in order to reduce overall execution time.
However, increasing parallelism can introduce crosstalk, which
is a significant source of errors in many quantum hardware platforms.
Therefore, there is a trade-off between achieving high parallelism
and minimizing error rates due to crosstalk.
The parallelism of a dynamic circuit is calculated as:
\begin{equation}
    P = (\frac{O_\text{total}}{D_\text{total}} - 1) \frac{1}{n - 1}
\end{equation}
where $n$ is the total number of qubits, $O_\text{total}$ is the total number of operations,
and $D_\text{total}$ is the total circuit depth.
\section{Benchmark Suite}
\label{sec:benchmark_suite}

\begin{figure*}
    \centering
    \includegraphics[width=0.9\textwidth]{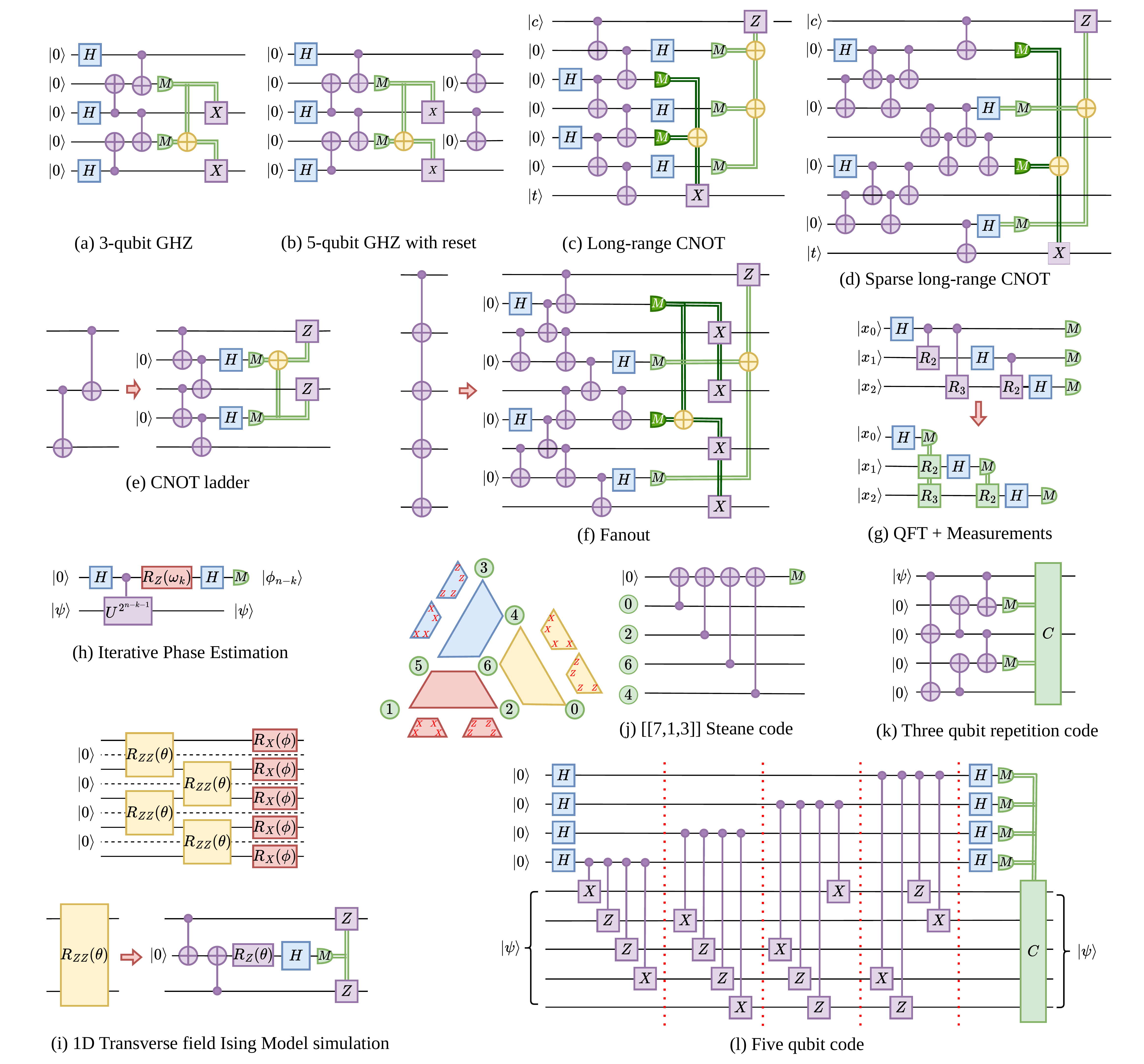}

    \caption{
    Dynamic circuit benchmark suite in \textit{dynamarq}.
    }
    \label{fig:circuits}
\end{figure*}

In this section, we introduce various dynamic circuit applications spanning state preparation, gate implementation, algorithms, and QEC. For each application, we propose a corresponding fidelity score that scales to large qubit counts and is normalized to $[0, 1]$ as described below.

\subsection{GHZ State Preparation}

The GHZ state is a highly entangled multi-qubit state that dynamic circuits can prepare at constant depth using ancilla qubits, compared to linear depth in unitary circuits \cite{baumer2024efficient}. As shown in Figure \ref{fig:circuits}(a), we use 5 qubits (3 data and 2 ancilla) to prepare a 3-qubit GHZ state. By applying a reset after each MCM, the ancilla qubits can be reused as system qubits, enabling preparation of a 5-qubit GHZ state, as shown in Figure~\ref{fig:circuits}(b).

\noindent\underline{\textit{Fidelity score}}: We use the Hellinger fidelity \cite{Hellinger} to compare the output probability distribution obtained from hardware with the ideal $n$-qubit GHZ state, represented by
\begin{equation}
 |\psi\rangle_{GHZ} =
    \frac{|0\rangle ^{\otimes n} + |1\rangle^{\otimes n}}{\sqrt 2}
\end{equation}

\subsection{Clifford Gates}
Recently, multi-qubit Clifford gates have been shown to be implementable using constant-depth dynamic circuits~\cite{baumer2024efficient, baumer2025measurement}. Our benchmarking suite includes the following constant-depth dynamic circuit-based Clifford gates: long-range CNOT,
CNOT ladder and the Fanout gate. 

\noindent\underline{\textit{Fidelity score}}: 

We adapt Direct fidelity estimation (DFE) \cite{flammia2011direct} to calculate the fidelity of a dynamic circuit implementing a Clifford operation $\mathcal{C}$ on $n$ data qubits and $a$ ancilla qubits, because it significantly reduces the exponential overhead of full quantum process tomography by requiring only a constant number of observable measurements. The process is as follows:
(1) We sample a uniformly random $n$-letter Pauli string $\mathcal{P}=P_1 P_2 \cdots P_n$ and prepare the stabilizer state $|\phi_1^{+}\rangle\otimes|\phi_2^{+}\rangle\otimes\cdots\otimes|\phi_n^{+}\rangle\otimes|0\rangle^{\otimes a}$, where $|\phi_i^+\rangle$ is the $+1$ eigenstate of $P_i$.
(2) We propagate the operator $\mathcal{P}$ through the Clifford $\mathcal{C}$ via stabilizer simulation \cite{aaronson2004improved}, obtaining $\mathcal{P}'=\mathcal{C}\mathcal{P}\mathcal{C}^\dagger$.
(3) We measure the $n$ data qubits in the $\mathcal{P}'$ basis by applying a final layer of single-qubit gates that converts $\mathcal{P}'$ to an operator $\mathcal{M}$ consisting of only $I$ and $Z$ Paulis on the data qubits and only $I$ on the ancilla qubits.
The fidelity is computed by averaging over $k = 30$ repetitions of the Pauli string sampling process. For each sample, every shot returns a bit string $b$ associated with the $+1$ or $-1$ eigenstate of $\mathcal{M}$, and we estimate $\langle \mathcal{M} \rangle$ by taking the expectation of the eigenvalue over $N$ shots.

\subsubsection{Long-range CNOT}

Implementing long-range CNOT gates on superconducting hardware is challenging due to limited qubit connectivity, typically requiring SWAP gates to bring qubits into interaction range, which increases circuit depth. Dynamic circuits offer a constant-depth implementation, where all qubits between the control and target are used as ancillas, with MCMs and feed-forward operations applied~\cite{baumer2024efficient}, as illustrated in Figure \ref{fig:circuits}(c). However, measuring all intermediate ancilla qubits limits their reusability and introduces additional errors. To address this, an alternative protocol has been proposed \cite{baumer2025measurement}, which we refer to as the ``sparse long-range CNOT''. It reduces the number of required ancillas and MCMs with only a modest increase in circuit depth, while still maintaining a constant depth, as shown in Figure \ref{fig:circuits}(d).

\subsubsection{CNOT ladder}

CNOT ladders are configurations of CNOT gates arranged in a ladder-like structure, used to generate multi-qubit entanglement and serve as key components in algorithms such as Trotterization for Hamiltonian simulation. Their unitary depth grows linearly with qubit count, but dynamic circuits with ancilla qubits enable constant-depth implementation on linear connectivity hardware \cite{baumer2025measurement}, as shown in Figure \ref{fig:circuits}(e).

\subsubsection{Fanout}
The quantum fanout gate on $n$ qubits takes the first qubit as the control and the remaining $n-1$ qubits as targets. The quantum fanout gate is known for its power in performing quantum arithmetic operators and reducing the depth of QFT~\cite{hoyer2005quantum}.
It can be realized using two layers of CNOT ladders,
thus its depth also grows linearly with the number of qubits.
We can also implement it using a dynamic circuit with constant depth
\cite{baumer2025measurement} as shown in Figure \ref{fig:circuits}(f).

\subsection{Quantum Algorithms}

We collect several quantum algorithms in our benchmarking suite detailed below; however, unlike SupermarQ~\cite{supermarq}, we exclude variational quantum algorithms such as QAOA~\cite{farhi2014quantum} and VQE~\cite{peruzzo2014variational}, as determining optimal parameters requires classical simulation or a prohibitively large number of quantum-classical iterations on hardware.

\subsubsection{QFT+Measurements}
Quantum Fourier Transform (QFT) is a fundamental primitive
that serves as a subroutine in many important algorithms,
like Shor's algorithm \cite{shor1999polynomial}
and Quantum Phase Estimation (QPE)~\cite{kitaev1995quantum}.
Standard QFT circuits require all-to-all connectivity, posing challenges
for quantum hardware with limited physical connectivity.
However, when QFT is followed by measurements,
as in QPE,
the deferred measurement principle \cite{baumer2024quantum} allows each controlled-rotation gate to be replaced by a classically controlled single-qubit rotation, effectively relaxing hardware connectivity requirements, as shown in Figure~\ref{fig:circuits}(g).
Since MCMs and feed-forward operations can introduce significant delay and
decoherence errors, we propose a hybrid approach, which we call ``partial QFT + M", that defers measurements for only a subset of qubits while the remaining qubits undergo the standard QFT procedure. This benchmark evaluates the trade-off between the errors introduced by MCMs and feed-forward operations and the benefits of alleviating hardware connectivity constraints,
specifically, the reduction of SWAP overhead. For simplicity, we assign half of the qubits to the dynamic QFT + M scheme and the other half to the standard QFT.

\noindent\underline{\textit{Fidelity score}}: Since QFT is not efficiently simulable, directly comparing the output distribution with the ideal result using statevector simulation or Hellinger fidelity would incur an exponential overhead. 
To address this issue, we adapt the fidelity calculation method
from \cite{qedc21}.
We select a random binary bitstring $s$, prepare all qubits in an equal superposition $H^{\otimes n}|0\rangle^{\otimes n}$, and apply single-qubit phase rotations to encode $s$ in the Fourier basis as $QFT|s\rangle$, followed by the inverse $QFT^{-1}$. The ideal output is then a delta function peaked at $s$, enabling the use of Hellinger fidelity to compare the output distribution against the ideal result. Note that, in adapting this method, we implement both a dynamic inverse QFT and a partial dynamic inverse QFT. We select three randomly sampled bitstrings $s$ and calculate the average as the fidelity score, following the same setting as \cite{qedc21}.

\subsubsection{Iterative Phase Estimation (IPE)}

IPE~\cite{o2010iterative, dobsicek2007arbitrary} is a space-efficient variant of QPE that estimates the phase $\phi \in [0,1)$, defined by the eigenvalue $e^{2\pi i \phi}$ of a unitary $U$ acting on eigenstate $|\psi\rangle$, using a single ancilla qubit with classical feed-forward instead of the large ancilla register required by standard QPE (Figure~\ref{fig:circuits}(h)). Writing $\phi = \sum_{j=1}^{n} \phi_j 2^{-j}$ in binary, IPE extracts each bit sequentially from $\phi_n$ to $\phi_1$. At each step $k$, assuming $\phi_{n-k+1}, \ldots, \phi_n$ are known: (1) prepare the ancilla in $|+\rangle$; (2) apply controlled-$U^{2^{n-k-1}}$, introducing a phase kickback; (3) apply a correction rotation $R_z(\omega_k)$ with $\omega_k = -2\pi\sum_{j=1}^{k}\phi_{n-k+j} 2^{-j-1}$, using previously measured bits to isolate $\phi_{n-k}$; and (4) measure the ancilla in the $X$ basis to determine $\phi_{n-k}$, feeding the result forward into the next iteration. We benchmark IPE using the phase gate $U_\theta = |0\rangle\langle 0| + e^{2\pi i\theta}|1\rangle\langle 1|$ with eigenstate $|1\rangle$ and $\theta \in [0,1)$.

\noindent{\underline{\textit{Fidelity score}}}:
Since $\theta$ is chosen and known exactly, the ideal IPE output is the $m$-bit binary representation of $\theta$, i.e., the bitstring $\theta_1\theta_2\ldots\theta_m$, occurring with probability 1. We define the fidelity score as the Hellinger fidelity between the measured output distribution and this ideal distribution.

\subsubsection{Hamiltonian simulation}

Hamiltonian simulation is one of the most promising fields for demonstrating quantum advantage, with broad applications in quantum chemistry and condensed matter physics. We focus on the 1D Transverse Field Ising Model (TFIM) in our benchmarking suite, because it can be efficiently simulated classically using Matrix Product State (MPS)-based methods~\cite{mps1d} and requires only linear connectivity, making it directly applicable to quantum hardware without SWAP gates.
Without periodic boundary conditions, the Hamiltonian is :
\begin{equation}
    H = -J \sum_{i=1}^{n-1}{\sigma_z^{i}\sigma_z^{i+1}} 
     - h \sum_{i=1}^{n}{\sigma_x^{i}}
\end{equation}
The Trotterized circuit for $\exp(-i \delta t H)$ consists of alternating layers of $R_{ZZ} = e^{-i\theta \sigma_z \otimes \sigma_z}$ gates between adjacent qubits and $R_X = e^{-i\phi \sigma_x}$ single-qubit gates. We implement this by interleaving $n-1$ ancilla qubits among the $n$ data qubits, realizing each two-qubit gate via dynamic circuits\cite{javadi2024quantum}. The full set of two-qubit gates is implemented using two layers of dynamic circuits, as illustrated in Figure \ref{fig:circuits}(i).

\noindent\underline{\textit{Fidelity score}}:
We simulate the circuit without
ancillas using Qiskit AerSimulator's \cite{qiskit} MPS simulator
to get the ideal average magnetization $\langle M_z \rangle_{ideal}$,
where $M_z$ is defined as
\begin{equation}
    M_z = \frac{1}{n}\sum_{i=1}^{n} \sigma_z^i
\end{equation}
The fidelity score is the relative error in average magnetization
$1 - \frac{|\langle M_z \rangle_{ideal} - \langle M_z \rangle_{observed}|}
    {|2 \langle M_z \rangle_{ideal}|}$.

\subsection{Quantum Error Correction} \label{sec:ecc}

Quantum error correction (QEC)~\cite{brun2019quantum, roffe2019quantum} is a primary driver for MCM capabilities, as it inherently requires syndrome measurement and feed-forward correction. However, current hardware supports only small-scale demonstrations, such as memory experiments \cite{surfacecodememory}, due to high physical error rates, limited real-time decoding capabilities, or limited qubit count.
While the Quantinuum/Guppy stack supports real-time decoding owing to longer coherence times~\cite{ransford2025helios}, IBM/Qiskit does not, precluding the benchmarking of more complex codes such as surface codes~\cite{fowler2012surface} or qLDPC codes~\cite{tillich2013quantum}. We therefore include three rudimentary error correction codes, specifically the bit-flip repetition code, the five-qubit code, and the Steane code, implemented with lookup-table decoders.
Since we only benchmark one logical qubit memory, our syndrome extraction circuits are not fault-tolerant.
Our benchmarks consist of state preparation, encoding, one round of syndrome measurement, and a final logical operator measurement.

\noindent{\underline{\textit{Fidelity score}}}:
The fidelity score is defined as one minus the logical error rate, i.e., the fraction of shots in which a logical error is detected.
States outside the code space when performing logical operator measurement are counted as logical errors.

\noindent{\underline{\textit{Branch probability}}}:
We estimate the branch probabilities by modeling the circuit-level noise
including one-qubit (idling) error, two-qubit gate errors and MCM errors. 
Let $p$ be the median two-qubit gate error rate, $s$ be the median single-qubit (idling) error rate
and $m$ be the median MCM error rate.
We assume $1-p,1-m,1-s\approx 1$, the second order terms $s^2,p^2,m^2,sp,pm,sm\approx 0$, and $s << p$.
Each QEC circuit contains one ancilla qubit per stabilizer measurement. The probability of a non-zero syndrome on a particular ancilla measuring a weight-$k$ stabilizer is approximated using three contributions: (1) exactly one of the $k$ two-qubit gates fails; (2) one MCM fails; (3) an idling error at some point during the circuit. This gives: $k p (1-p)^{k-1} (1-m)(1-s) + (1-p)^k m (1-s) + (1-p)^k(1-m)s \approx k p + m + s$.
Single-qubit idling errors on a data qubit that flip multiple ancillas are ignored because $s$ is typically
an order of magnitude smaller than $p$.
For IBM quantum hardware, $p \sim 10^{-3}$, $s \sim 10^{-4}$, and $m \sim 5\times10^{-3}$, yielding a branch probability for corrections of approximately 1 to 3\%. For the Quantinuum Helios emulator~\cite{ransford2025helios}, the error rates are significantly lower: $p \approx 8\times 10^{-4}$, $s \approx 2.5\times10^{-5}$, and $m \approx 10^{-6}$, yielding branch probabilities of approximately 0.1 to 0.3\%.

These estimates align with our hardware observations, except for the Five-Qubit Code (Section \ref{sec:fqc}), which suffers from SWAP overhead when compiled to IBM hardware. To obtain a more accurate branch probability estimation for circuits with routing overhead, we conduct noise simulations on the transpiled circuits using Stim~\cite{gidney2021stim} with the same noise sources as above.

\subsubsection{Bit flip codes}

The $n$-qubit bit-flip code encodes $|0\rangle \rightarrow |0\rangle^{\otimes n}$ and $|1\rangle \rightarrow |1\rangle^{\otimes n}$, correcting up to $\frac{n-1}{2}$ $X$ errors. We restrict our benchmarks to $n \in \{3, 5\}$, as the complexity of the correction circuit grows exponentially with $n$. We encode the $|1\rangle$ and $|+\rangle$ states, with $n-1$ ancilla qubits interleaved among the $n$ data qubits for parity checks, as shown in Figure~\ref{fig:circuits}(k). 
The probability of a single bit flip on a parity check (ancilla) qubit is approximately $2p+m+s$.

\subsubsection{Five-qubit code}
\label{sec:fqc}
The [[5,1,3]]  error correcting code \cite{gottesman2009introduction}
is a stabilizer code \cite{gottesman1997stabilizer}
with stabilizers $XZZXI$ and its three cyclic permutations (Figure \ref{fig:circuits}(l)).
Its logical operators are $\bar{X}=XXXXX$ and $\bar{Z}=ZZZZZ$.
It detects any single qubit $X$, $Y$ or $Z$ error, making it
robust to more general noise models than the bit flip noise.
We encode the $|0\rangle$ and the $|1\rangle$ states. 
There are 16 possible syndromes including the all-zero syndrome, with correction branch probabilities of approximately $4p + m + s$.
Our model initially predicts a branch probability of approximately 2 to 3\%. However, the syndrome extraction circuit is deep due to long range CNOT gates that require SWAP gates during transpilation, which introduce significant noise. To account for this, we use Stim \cite{gidney2021stim} to perform branch probability estimation directly on the \textbf{transpiled} circuit, and we obtain a uniform probability of $\frac{1}{16} = 6.25\%$ for each branch, yielding results that are consistent with our hardware observations.

For the Quantinuum Helios emulator, we still use approximated branch probability based on hardware noise model due to its all-to-all connectivity and lower noise rate.

\subsubsection{Steane code}
The [[7,1,3]] Steane code \cite{PhysRevLett.77.793} is a distance 3 self-dual CSS code with the parity check matrix given by the [7,4,3] Hamming code.
It has been recently used to perform logical operations fault tolerantly on Quantinuum's Helios trapped ion system \cite{perlin2026fault}.
It can correct any 1-weight Pauli error. We encode $|1\rangle$ and $|+\rangle$ states.
Figure \ref{fig:circuits}(j) shows the six plaquettes corresponding to the six stabilizer checks, each of which is measured by an ancilla qubit via a circuit like the one shown in the same figure.
We find that the $X$ and $Z$ correction branch probabilities are approximately $4p+m+s$.
\section{Evaluation}
 \label{sec:evaluation}
\begin{figure}
    \centering
    \includegraphics[width=0.25\textwidth]{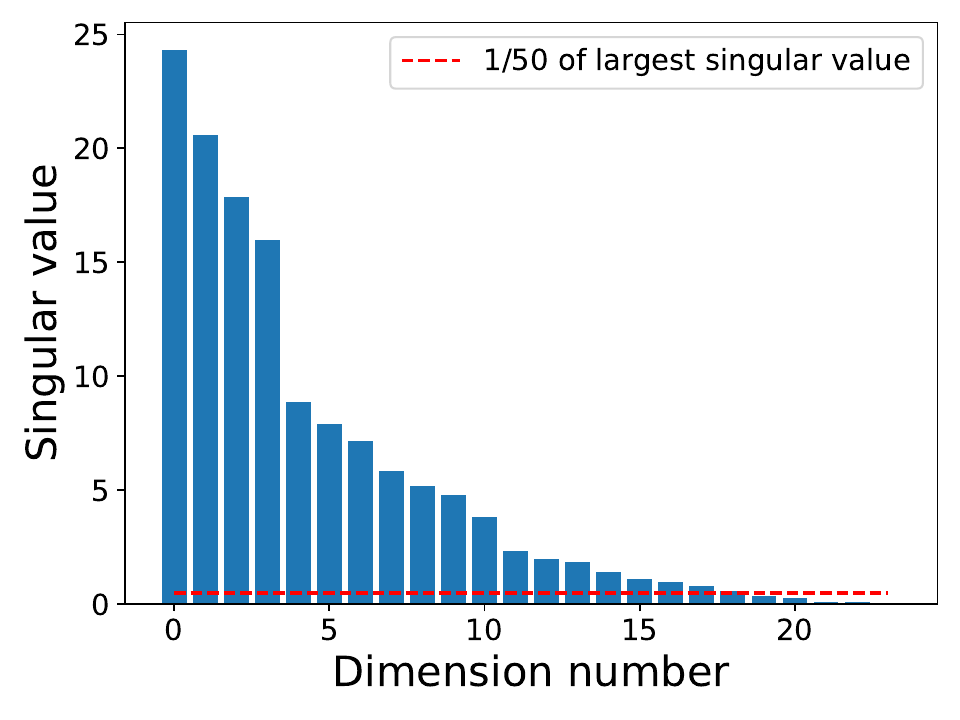}
    \caption{
   Histogram of singular values from PCA on the feature matrix for IBM Pittsburgh (no DD). Results are consistent across all IBM backends with and without DD.
    }
    \label{fig:pca}
\end{figure}

\textbf{Benchmark selection.}We collect circuits spanning a range of qubit counts for each application, with details listed in Table~\ref{table:benchmark_list}.
For IBM Quantum hardware, the upper qubit count is determined by when the fidelity score approaches zero,
except for QFT and IPE benchmarks, where it is capped at 20 qubits, beyond which rotation angles become negligibly small,
and QEC codes where the range is code-dependent and is limited by other factors like decoder complexity.
Note that QEC codes do not require arbitrary scalability, as logical error rates drop exponentially with distance, making very large-distance codes unnecessary in practice. Thus, the scalability requirement for QEC benchmarks can be relaxed as long as \textit{dynamarq} supports sufficiently large codes.
For the Quantinuum Helios-1E emulator, the range is capped at the maximum supported qubit count of 25.
Long-range CNOTs (sparse) are excluded from Quantinuum machine due to its all-to-all connectivity,
and (partial) QFT and IPE are excluded due to limitations in the Guppy library.

\textbf{Quantum hardware/emulator.} We run all circuits on two IBM superconducting quantum processors, IBM Kingston (Heron r2) and IBM Pittsburgh (Heron r3), and the Quantinuum Helios-1E trapped-ion emulator (with a very accurate noise model compared to real quantum hardware \cite{montanez2025evaluating}), with 4096 shots per circuit. For IBM quantum hardware, we utilize IBM's newly introduced MCM feature (shorter duration than the final measurement) and additionally explore dynamical decoupling (DD)~\cite{viola1998dynamical} to suppress idling errors on data qubits during MCM and feed-forward operations.

\textbf{Circuit features.} We compute circuit features for all benchmarks
\textbf{without transpilation} to ensure hardware agnosticism, and transpile the circuits only prior to hardware execution.
An end user may evaluate features for transpiled circuits if they choose.
Some features have variants depending on whether MCM, reset, and feed-forward operations are included, as detailed in Table~\ref{table:feature_metrics}.
For the IBM data, we perform principal component analysis (PCA) to assess linear independence among features, as shown in Figure~\ref{fig:pca}. At a cutoff of $1/50\times$ the largest singular value, we find 17 dominant directions, confirming that all the features contribute meaningfully and are retained in the analysis.
PCA is not performed for the Quantinuum emulator due to the limited number of available benchmarks.

\textbf{Statistical modeling.}
We train regularized linear regression models using circuit features and fidelity scores obtained from hardware experiments.
We evaluate our models in three ways: (1) computing $R^2$ scores on the full dataset to characterize the correlation between circuit features and fidelity scores; (2) splitting the dataset into train/test sets to evaluate the model's ability to predict fidelity scores of unseen benchmarks; and (3) evaluating the transferability of model parameters across different quantum hardware backends and calibration cycles.
While our regression combines benchmarks with different fidelity metrics, this heterogeneity is intentional: each metric is application-specific, normalized, and efficiently computable for large circuits, providing a common scale across different metric definitions. Rather than limiting machine performance to a single aggregate $R^2$ score, \textit{dynamarq} outputs a per-application fidelity profile, and users may restrict the model to any subset of benchmarks relevant to their application.

\textbf{Baselines.} \textit{dynamarq} provides both normalized and unnormalized features. For normalized features, we compare against SupermarQ~\cite{supermarq}, which proposes six normalized features: communication, critical depth, entanglement, liveness, measurement, and parallelism. For unnormalized features, we compare against several commonly used unnormalized circuit features~\cite{qedc21, murali2019full}, including circuit depth, number of qubits, and number of two-qubit gates
(referred to as SOTA-Unnorm). Note that none of these baselines account for branching information, MCM, or feed-forward classical control in dynamic circuits.

\begin{table}[htbp]
  \caption{List of benchmarks with range of total qubit counts (system + ancilla) for different backends.}
  \begin{center}
  \resizebox{0.7\columnwidth}{!}{
  \begin{tabular}{|c|c|c|}
    \hline
    \textbf{Benchmark} & \textbf{Backend} & \textbf{Qubits}\\
    \hline
    \multirow{2}{*}{GHZ} & IBM & 3-59 \\
    \cline{2-3}
    & Helios-1E & 3-19 \\
    \hline
    \multirow{2}{*}{GHZ (reset)} & IBM & 3-59 \\
    \cline{2-3}
    & Helios-1E & 3-25 \\
    \hline
    \multirow{2}{*}{CNOT Ladder} & IBM & 3-59 \\
    \cline{2-3}
    & Helios-1E & 3-19 \\
    \hline
    \multirow{2}{*}{Fanout} & IBM & 5-61 \\
    \cline{2-3}
    & Helios-1E & 5-21 \\
    \hline
    Long range CNOT & IBM & 4-32 \\
    \hline
    Long range CNOT (sparse) & IBM & 5-61 \\
    \hline
    QFT & IBM & 2-20 \\
    \hline
    Partial QFT & IBM & 2-20 \\
    \hline
    IPE & IBM & 2 \\
    \hline
    \multirow{2}{*}{TFIM Simulation} & IBM & 5-59 \\
    \cline{2-3}
    & Helios-1E & 5-19 \\
    \hline
    Repetition Code & IBM/Helios-1E & 5, 9 \\
    \hline
    Five qubit code & IBM/Helios-1E & 11 \\
    \hline
    Steane code & IBM/Helios-1E & 14 \\
    \hline
  \end{tabular}
  }
  \end{center}
  \label{table:benchmark_list}
\end{table}

\begin{table}[htbp]
  \caption{List of \textit{dynamarq} circuit features, whether they are \underline{U}n/\underline{N}ormalized and their descriptions.}
  \begin{center}
   \resizebox{0.78\columnwidth}{!}{
  \begin{tabular}{|c|c|c|}
    \hline
    \textbf{\#} & U/N & \textbf{Feature description}\\
    \hline
    0 & U & Depth (w/o FF operations) \\
    \hline
    1 & U & Depth (with FF operations) \\
    \hline
    2 & U & \# Operations (w/o meas/resets/FFs) \\
    \hline
    3 & U & \# Operations (w/o FFs, but with meas/resets) \\
    \hline
    4 & U & \# Operations (with meas, resets and FFs) \\
    \hline
    5 & U & \# System qubits (carrying information) \\
    \hline
    6 & U & \# Total qubits (data + ancilla qubits) \\
    \hline
    7 & N & Liveness (w/o FF) \\
    \hline
    8 & N & Liveness (with FF) \\
    \hline
    9 & N & System-qubit ratio \\
    \hline
    10 & N & Critical path (only quantum operations) \\
    \hline
    11 & N & Critical path (quantum + classical operations) \\
    \hline
    12 & N & Dynamic depth ratio (w/o FF) \\
    \hline
    13 & N & Dynamic depth ratio (with FF) \\
    \hline
    14 & N & Parallelism (w/o FF operations) \\
    \hline
    15 & N & Parallelism (with FF operations) \\ 
    \hline
    16 & N & Quantum communication \\
    \hline
    17 & N & Quantum + Classical Communication \\
    \hline
    18 & N & Quantum entanglement (w/o meas/resets/FFs) \\
    \hline
    19 & N & Quantum entanglement (w/o FFs, with meas/resets) \\
    \hline
    20 & N & Quantum entanglement (with FFs/meas/resets) \\
    \hline
    21 & N & Quantum+FF corr. (w/o meas/resets/FFs) \\
    \hline
    22 & N & Quantum+FF corr. (w/o FFs, with meas/resets) \\
    \hline
    23 & N & Quantum+FF corr. (with FFs/meas/resets) \\
    \hline
  \end{tabular}
    }
  \end{center}

  \label{table:feature_metrics}
\end{table}

\subsection{Statistical Modeling Results}

\subsubsection{Feature-fidelity correlation analysis.}
\label{sec:regression}

\begin{figure*}[htpb]
\centering
\includegraphics[width=\textwidth]{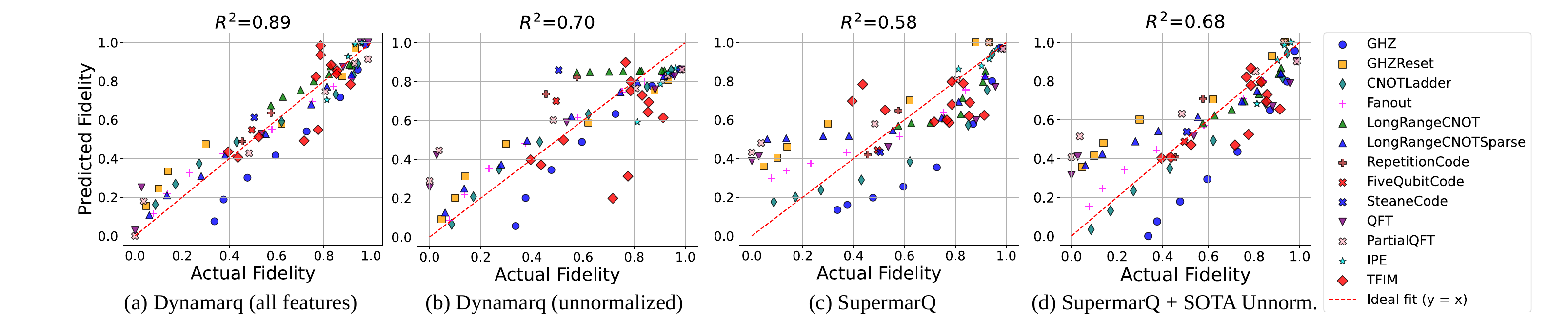}
\caption{
Linear regression results for the correlation between circuit features and fidelity scores on IBM Pittsburgh under different feature settings. Higher $R^2$ value indicates greater predictive capability.
}

\label{fig:regression}
\end{figure*}

We train a linear regression model on the entire dataset to analyze the correlations between circuit features and fidelity scores under four settings: (1) all \textit{dynamarq} features (both unnormalized and normalized), (2) unnormalized \textit{dynamarq} features only, (3) SupermarQ features only, and (4) SupermarQ + SOTA-Unnorm, which include both normalized and unnormalized features.
The results for IBM Pittsburgh are shown in Figure~\ref{fig:regression}. Using all \textit{dynamarq} features, we obtain the highest $R^2$ of $0.89$, indicating that our features are strongly correlated with fidelity scores. Using only the unnormalized \textit{dynamarq} features only yields a moderate $R^2$ of $0.70$, demonstrating the importance of using both unnormalized and normalized features proposed in Section \ref{sec:features}.
In contrast, using SupermarQ features drops the score to 0.58, and SupermarQ + SOTA-Unnorm does not lead to a substantial improvement. This suggests that features derived from unitary circuits cannot effectively capture the characteristics of dynamic circuits, particularly in the low-fidelity regime, which becomes increasingly prevalent as circuit depth, qubit count, and gate count grow. Overall, our approach achieves an $R^2$ improvement of 53.4\% over SupermarQ and 30.8\% over SupermarQ + SOTA-Unnorm, and similar results are observed on IBM Kingston. For the Quantinuum Helios emulator, \textit{dynamarq} achieves an $R^2$ score of 0.95, which is $2.8\times$ higher than SupermarQ and 5.5\% higher than SupermarQ + SOTA-Unnorm. We additionally evaluate our model under DD for IBM quantum hardware, which introduces single-qubit gates that slightly modify feature values while generally improving fidelity scores. Our model achieves an even higher $R^2$ with DD using \textit{dynamarq} features, demonstrating its robustness across both settings. The complete $R^2$ scores using different features settings for all hardware configurations with and without DD are listed in Table~\ref{table:fitting_scores}.

\begin{table}[h]
\caption{$R^2$ scores for data fitting on different hardware configurations using various feature settings:
  (a) All \textit{dynamarq} features
  (b) Only unnormalized \textit{dynamarq} features
  (c) SupermarQ features
  (d) SupermarQ + SOTA Unnorm. features.}
\centering
       \resizebox{0.7\columnwidth}{!}{

\begin{tabular}{|l|l|l|l|l|}
\hline
\multicolumn{1}{|c|}{\multirow{2}{*}{Hardware configuration}} & 
    \multicolumn{4}{c|}{$R^2$ score} \\ \cline{2-5} 
\multicolumn{1}{|c|}{}   & (a)  & (b)  & (c)  & (d)  \\ \hline
IBM Kingston (w/o DD)    & \textbf{0.86} & 0.70 & 0.62 & 0.67 \\ \hline
IBM Kingston (with DD)   & \textbf{0.90} & 0.65 & 0.39 & 0.54 \\ \hline
IBM Pittsburgh (w/o DD)  & \textbf{0.89} & 0.70 & 0.58 & 0.68 \\ \hline
IBM Pittsburgh (with DD) & \textbf{0.91} & 0.57 & 0.39 & 0.51 \\ \hline
Quantinuum Helios-1E     & \textbf{0.95} & 0.89 & 0.34 & 0.90 \\ \hline 
\end{tabular}
}
\label{table:fitting_scores}
\end{table}

\subsubsection{Individual feature-fidelity correlation analysis.}
\label{sec:correlation}

\begin{figure*}
\centering
\includegraphics[width=0.9\textwidth]{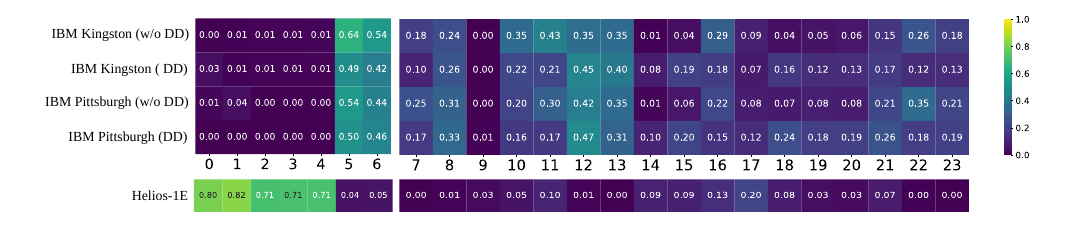}
\caption{
  Individual correlations between features and fidelity scores across quantum backends.
  Rows correspond to backends and the columns are features in Table
  \ref{table:feature_metrics} in order, where indices 0--6 are unnormalized and 7--23 are normalized features.
}
\label{fig:dynamarq_heatmap}
\end{figure*}

Figure~\ref{fig:dynamarq_heatmap} shows the correlations between individual features (listed in Table~\ref{table:feature_metrics}) and fidelity scores across five settings: (1) IBM Kingston without DD, (2) IBM Kingston with DD, (3) IBM Pittsburgh without DD, (4) IBM Pittsburgh with DD, and (5) Quantinuum Helios-1E. The correlations vary across hardware modalities but remain consistent with and without DD. For IBM superconducting hardware, whether DD is applied, the three most strongly correlated features are the number of qubits, the dynamic depth ratio, and liveness. For Quantinuum trapped-ion Helios-1E, the three most correlated features are circuit depth, number of operations, and quantum-classical communication.

\subsubsection{Fidelity prediction on unseen data.}
\label{sec:train_test_split}

To evaluate the ability of our models to predict fidelity on unseen circuits,
we perform 50 random 80/20\% train/test splits for each IBM quantum hardware configuration (with
and without DD), training on the train split and evaluating the $R^2$ score on the test split. 
The averaged $R^2$ results are shown in Table~\ref{table:prediction_scores}, where we achieve consistently high scores, all $\geq 0.72$, demonstrating strong predictive performance on unseen circuits.
We also evaluate generalization to unseen circuit structures by training on all benchmark types except one and testing on the held-out type. The results show that when the training set contains benchmarks with similar structures to the test type, the model generalizes well and achieves a high $R^2$ score. For example, testing on Fanout circuits yields an $R^2$ of $0.92$, as their structure is similar to CNOT Ladder circuits included in the training set. However, for completely novel structures with no similar counterpart in the training set, performance can degrade significantly. For instance, IPE circuits yield an $R^2$ below zero, indicating that the model fails to generalize to entirely unseen structures. For Quantinuum Helios-1E, the limited qubit count results in insufficient data points to meaningfully train a prediction model.

\begin{table}[htbp]
  \caption{$R^2$ scores for predicting fidelities on unseen data.}
  
  \begin{center}
  \small
       \resizebox{0.9\columnwidth}{!}{

  \begin{tabular}{|c|c|c|}
    \hline
    \textbf{Training Set} & \textbf{Test Set} & \textbf{$R^2$}\\
    \hline
    IBM Kingston (w/o DD) 80\% & IBM Kingston (w/o) 20\% & 0.72 \\
    \hline
    IBM Kingston (DD) 80\% & IBM Kingston (DD) 20\% & 0.84 \\
    \hline
    IBM Pittsburgh (w/o) 80\% & IBM Pittsburgh (w/o) 20\% & 0.81 \\
    \hline
    IBM Pittsburgh (DD) 80\% & IBM Pittsburgh (DD) 20\% & 0.84 \\
    \hline
    Quantinuum Helios-1E 80\% & Helios-1E 20\% & $<$0 \\
    \hline
    IBM Pittsburgh (w/o DD) excl. Fanout & Fanout & 0.92 \\
    \hline
    IBM Pittsburgh (w/o DD) excl. CNOT Ladder & CNOT Ladder & 0.86 \\
    \hline
    IBM Pittsburgh (w/o DD) excl. IPE & IPE & $<$0 \\
    \hline
  \end{tabular}
  }
  \end{center}
  \label{table:prediction_scores}
\end{table}

\begin{table}[htbp]
  \caption{$R^2$ scores for predicting fidelities in cross-backend and cross-calibration cycle regimes.}
  \begin{center}
  \small
         \resizebox{0.85\columnwidth}{!}{

  \begin{tabular}{|c|c|c|}
    \hline
    \textbf{Training Set} & \textbf{Test Set} & \textbf{$R^2$}\\
    \hline
    IBM Kingston (w/o DD) & IBM Pittsburgh (w/o DD) & 0.72 \\
    \hline
    IBM Pittsburgh (w/o DD) & IBM Kingston (w/o DD) & 0.72 \\
    \hline
    IBM Kingston (DD) & IBM Pittsburgh (DD) & 0.79 \\
    \hline
    IBM Pittsburgh (DD) & IBM Kingston (DD) & 0.81 \\
    \hline
    IBM Pittsburgh (w/o DD) run 1 & IBM Pittsburgh (w/o DD) run 2 & 0.71 \\
    \hline
  \end{tabular}
  }
  \end{center}
  \label{table:cross_backend_scores}
\end{table}
\begin{figure}[htbp]
\centering
\includegraphics[width=0.45\textwidth]{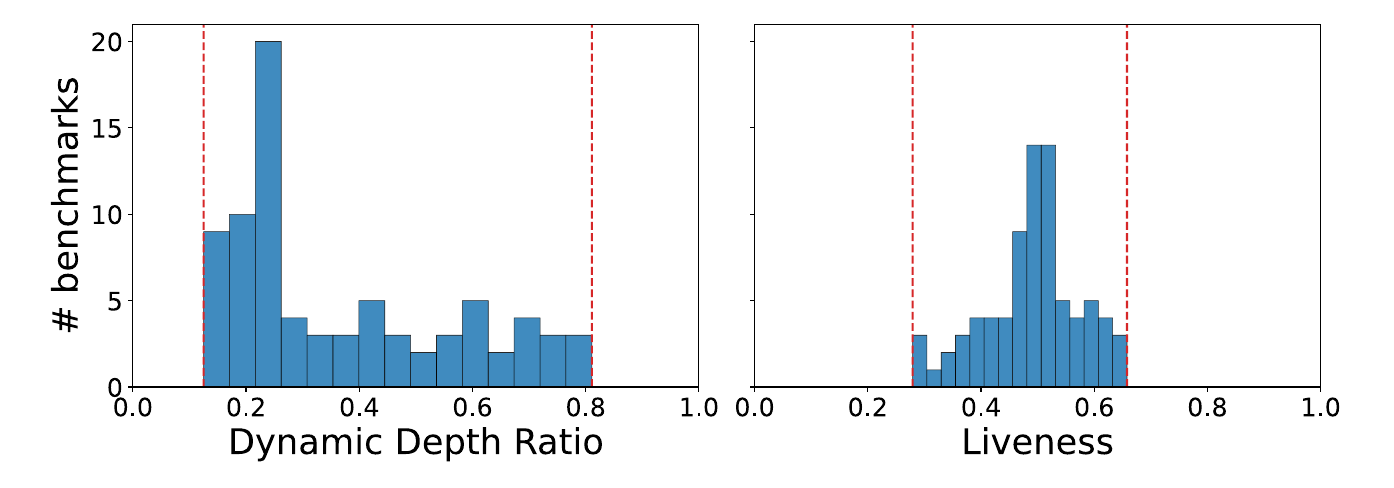}
\caption{
Coverage histograms of two normalized features most correlated with circuit fidelity executed on IBM Pittsburgh (w/o DD).
}
\label{fig:coverage}
\end{figure}

\subsubsection{Feature coverage analysis}
\label{sec:coverage}
Since \emph{dynamarq}'s prediction quality depends on whether a new circuit falls within the feature space covered by the training data, we provide a feature coverage analysis from two angles. First, we evaluate the feature space spanned by the benchmark points. As shown in Figure~\ref{fig:coverage}, we plot the two most correlated normalized features for IBM quantum hardware, and the histograms confirm that the benchmarks span a broad region of the feature space. Second, we provide a feature coverage checker based on a k-nearest neighbour (kNN) classifier. A user can run this check before trusting a predicted fidelity: if the new circuit's feature vector lies within the trained region, the prediction can be used directly; otherwise, the user should run the circuit on hardware, collect the fidelity data, and retrain the model. To evaluate the classifier, we use two protocols. In leave-one-family-out (LOFO) evaluation, we remove an entire circuit family (e.g., IPE) from training and treat each benchmark in that family as a test sample. For each such benchmark, we compare the classifier's output against whether the predicted and actual fidelities agree within a tolerance of 0.2. In leave-one-benchmark-out (LOBO) evaluation, only a single benchmark is held out, representing the more realistic use case. We achieve 71\% kNN classifier accuracy under LOFO and 77\% under LOBO on IBM Pittsburgh.

\begin{figure}[h]
\centering
\includegraphics[width=0.42\textwidth]{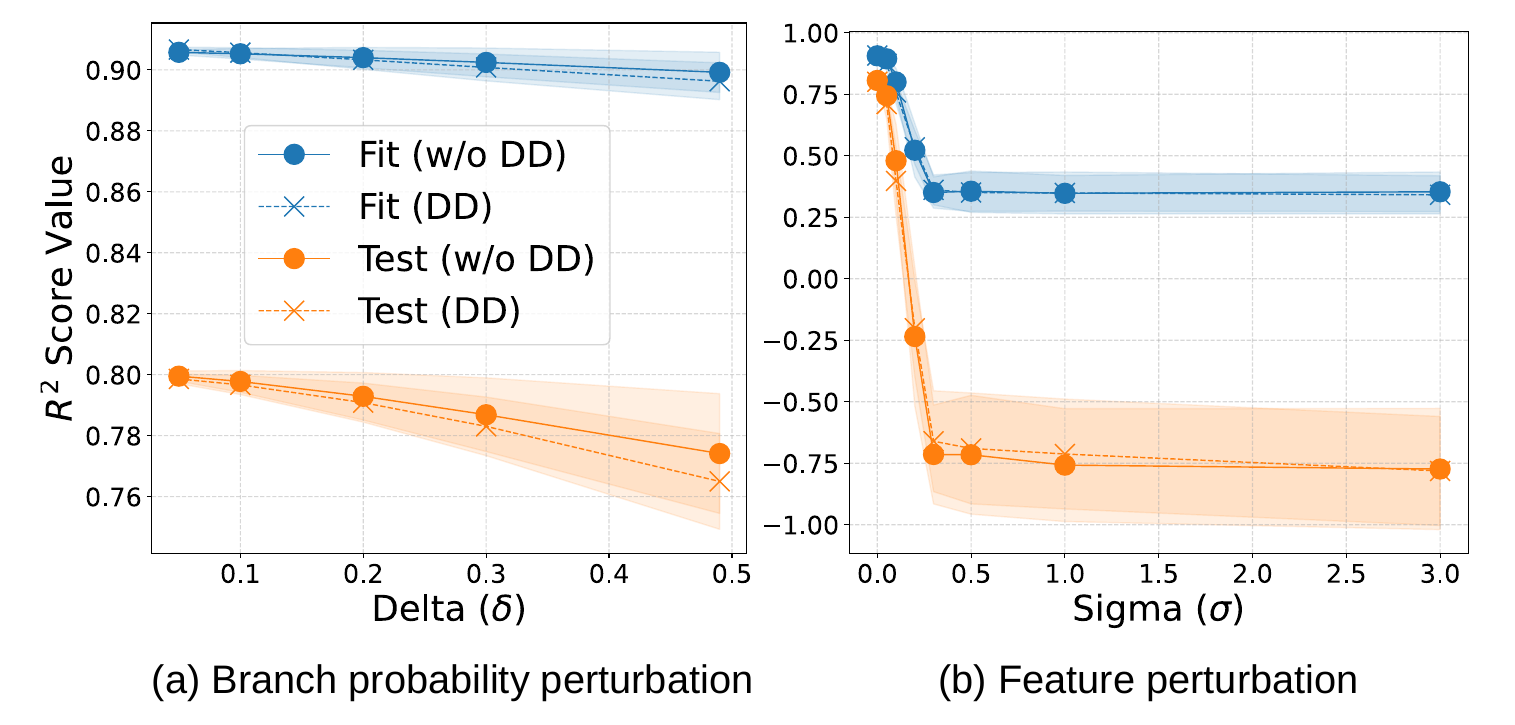}
\caption{
Sensitivity analysis under branch probability and feature perturbation on  IBM Pittsburgh (w/o DD).
}
\label{fig:sensitivity}
\end{figure}

\subsubsection{Sensitivity analysis}
\label{sec:sensitivity}

We study the effects of deviations from the uniform branch probability assumption of Section \ref{sec:prob}. We perturb the branch probability input to \textit{dynamarq}'s feature computation by randomly sampling values from 
$[0.5-\delta,0.5+\delta]$ independently for every qubit measurement outcome, and observe the effect on $R^2$ scores for both the data fitting experiment (Section \ref{sec:regression}) and the fidelity prediction experiment (Section \ref{sec:train_test_split}). We then study the effects of errors in feature estimation by adding i.i.d. Gaussian noise of strength $\sigma$ to every standardized feature and re-running the same experiments. The results are shown in Figure \ref{fig:sensitivity}. For branch probability perturbation, the $R^2$ scores degrade by only about 5\% in absolute value, demonstrating robustness to large deviations. For feature perturbation, however, we observe a steep decline in $R^2$ scores for both fitting and prediction, indicating that the model requires accurately estimated features to maintain good predictive performance.

\subsubsection{Parameter transferability}
\label{sec:parameter_transferability}

To evaluate the parameter transferability of our model across hardware backends, we treat the dataset from one machine (e.g., IBM Kingston) as the training set and another (e.g., IBM Pittsburgh) as the test set. The results are shown in Table~\ref{table:cross_backend_scores}, where we observe $R^2 \approx 0.72$--$0.81$, demonstrating that the model parameters are transferable across IBM quantum hardware backends and that fidelity scores on a new backend can be effectively predicted using parameters trained on another. Note that both IBM systems share similar Heron r2/r3 architecture, currently the only ones supporting optimized MCM and dynamic circuit capabilities (e.g., if-else instruction). We will extend this evaluation to more diverse architectures once they provide such support.
We further evaluate the transferability of our model across calibration cycles, as IBM quantum hardware is calibrated regularly. We treat the dataset obtained from one run of our benchmark suite as the training set and a run from a different day as the test set. We observe $R^2 \approx 0.71$, demonstrating that the model parameters are also transferable across calibration cycles.
\section{Potential applications of \textit{dynamarq}}
\label{sec:applications}
\textit{dynamarq} can be used for a variety of applications by a wide range of practitioners. Here we summarize three use cases. (1) \textbf{Hardware resource saving}: Before running a circuit on hardware, a user can first apply our kNN-based feature coverage checker (See Section~\ref{sec:coverage}) to determine whether the circuit falls within the trained distribution. If so, \textit{dynamarq} directly provides a reliable fidelity estimate without any hardware execution, saving costly quantum resources. When models are available for multiple platforms (e.g., IBM and Quantinuum), a user can predict fidelity on each platform, and choose the hardware with the highest predicted fidelity without executing the circuit on other machines. In this way, a user can choose the best hardware on which to run their circuit without the need for costly per-circuit benchmarking across multiple machines. (2) \textbf{Compiler-pass decisions}: As dynamic circuits become increasingly common, the same quantum algorithm can often be implemented using different dynamic circuit primitives (e.g., different repeat-until-success (RUS) constructions~\cite{paetznick2013repeat}). \textit{dynamarq} can estimate the fidelity of each candidate implementation based on its features, allowing a dynamic circuit compiler
\cite{niu2024acdcautomatedcompilationdynamic,10992761,quetschlich2023predicting}
to automatically select the best circuit design for the target hardware without exhaustive physical benchmarking.
(3) \textbf{Hardware-specific fidelity prediction}: Hardware-specific fidelity prediction is an active research area, with recent works training GNNs, LSTMs, and other ML models on combinations of circuit features and hardware calibration data to predict execution fidelity on specific machines~\cite{vadali2024quantum,mao2025q,saravanan2022machine,wang2022quest}.
These models (and compilers from (2)) rely on circuit features as a key input, and SupermarQ features are widely adopted for this purpose on static circuits.
However, no existing feature set captures the structural properties of dynamic circuits such as mid-circuit measurements, feed-forward operations, and branching behavior.
\textit{dynamarq} fills this gap by providing the first hardware-agnostic feature set designed specifically for dynamic circuits, making it a natural drop-in replacement for SupermarQ features in any hardware-specific fidelity prediction pipeline that needs to handle dynamic circuits.
\section{Conclusion and Future Work}
  \label{sec:conclusion}

In this work, we introduce \textit{dynamarq}, a hardware-agnostic, scalable benchmarking suite for dynamic quantum circuits. \textit{dynamarq} encompasses a diverse set of application-level benchmarks, circuit feature metrics for characterizing dynamic circuit structures, and scalable fidelity scores for evaluating hardware executions. Through statistical modeling, we identify correlations between circuit features and fidelity scores, and demonstrate the predictive power of our model on unseen circuits,
whose features are covered by the training distribution, as well as on unseen machines that are similar in architecture to the training data sample sources. Dynamic circuits are poised to play a central role in the future of quantum computing, particularly in fault-tolerant quantum computing (FTQC). Our future work will focus on two key areas: integrating advanced QEC codes with realistic hardware noise models for logical error prediction, and extending our dynamic feature analysis to support additional FTQC protocols like repeat-until-success\cite{paetznick2013repeat}, magic state distillation~\cite{bravyi2005universal}, and state injection~\cite{kim2026magic}.

\section*{Acknowledgment}
SS is supported by University of Central Florida ORCGS Doctoral Fellowship. EK, AM, WAdJ and CI were supported by the U.S. Department of Energy, Office of Science, Office of Advanced Scientific Computing Research under Contract No. DE-AC02-05CH11231, through the Accelerated Research in Quantum Computing Program, FAR-QC and MACH-Q. This research used resources of the Oak Ridge Leadership Computing Facility, a DOE Office of Science User Facility supported under Contract DE-AC05-00OR22725, and the National Energy Research Scientific Computing Center (NERSC), a Department of Energy Office of Science User Facility under Contract No. DE-AC02-05CH11231 using NERSC award DDR-ERCAP0034530.

\appendices

\section{Artifact Appendix}

\subsection{Artifact Details}
The artifacts contain code to reproduce all of the figures in the evaluation section.
\textit{dynamarq} \cite{dynamarq} is available as a pyPI package, installable via ``pip install dynamarq''
which includes the circuit generators and feature metric computation libraries.
The scripts to submit these jobs and to get the outputs from running these circuits
on quantum hardware/emulator, and the subsequent statistical analysis are all provided
as Jupyter notebooks at the following links :
\href{https://github.com/sumeetshirgure/dynamarq-execution}{github.com/sumeetshirgure/dynamarq-execution}

\begin{itemize}
    \item \textbf{Language} : Python
    \item \textbf{License} : MIT License
    \item \textbf{Software requirements} : Standard Mac/Linux/PC that can install python packages
            from pyPI and can run jupyter notebooks. All dependencies are freely available online.
    \item \textbf{Hardware requirements} : Minimal. Standard issue laptop with 2GB disk space and 4-8 GB RAM.
    \item \textbf{Runtime environments} : Standard python \texttt{venv} or \texttt{conda} environments.
    \item \textbf{Time needed} : 30 mins for running all Jupyter notebooks.
            A day or two to submit jobs to IBM Quantum/Quantinuum Nexus
            and get the results depending on their service queue.
            (4-6 hours of access credits to submit jobs to IBM Quantum.
            And about 2-3 hours of access credits for Quantinuum emulator.
            In case one can't access these systems, we also provide the outputs we received and
            their job IDs/metadata in the Zenodo link.)
    \item \textbf{Publicly available} : Yes
    \item \textbf{Artifact DOI} :
            \href{https://doi.org/10.5281/zenodo.21459804}{https://doi.org/10.5281/zenodo.21459804}
            The Zenodo link also contains the quantum computer/emulator outputs used to
            generate the plots, but due to space constraints these outputs are not tracked on GitHub.
\end{itemize}

\subsection{Instructions for Generating Plots}
Clone the GitHub repository and run the Jupyter notebooks in any order. The repository has a cache
of the output summary of the quantum hardware/emulator jobs submitted by the authors.
If one wants to re-run the circuits on quantum hardware/emulator
the helper scripts to submit the jobs (and track their job IDs) are also provided.
Each figure in the paper can be reproduced once the quantum computer outputs are available in the
respective locations that the Jupyter notebooks read from.

\section{Hardware fidelity data}

In this section we provide the hardware fidelity scores observed during hardware runs.

\begin{figure*}
    \centering
    \includegraphics[width=\textwidth]{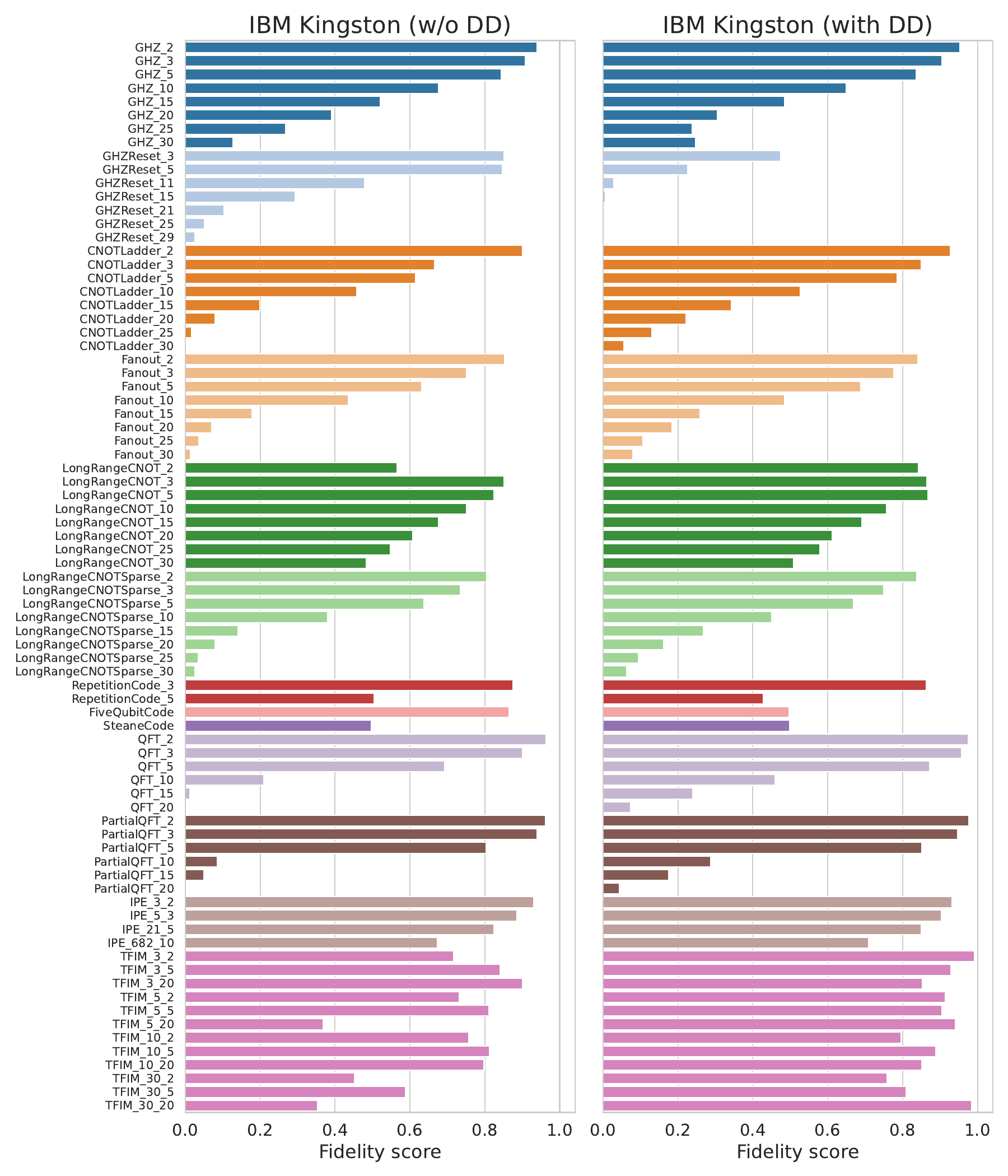}
    \caption{
    Fidelity scores on IBM Kingston using benchmarks in \textit{dynamarq}.
    }
\end{figure*}

\begin{figure*}
    \centering
    \includegraphics[width=\textwidth]{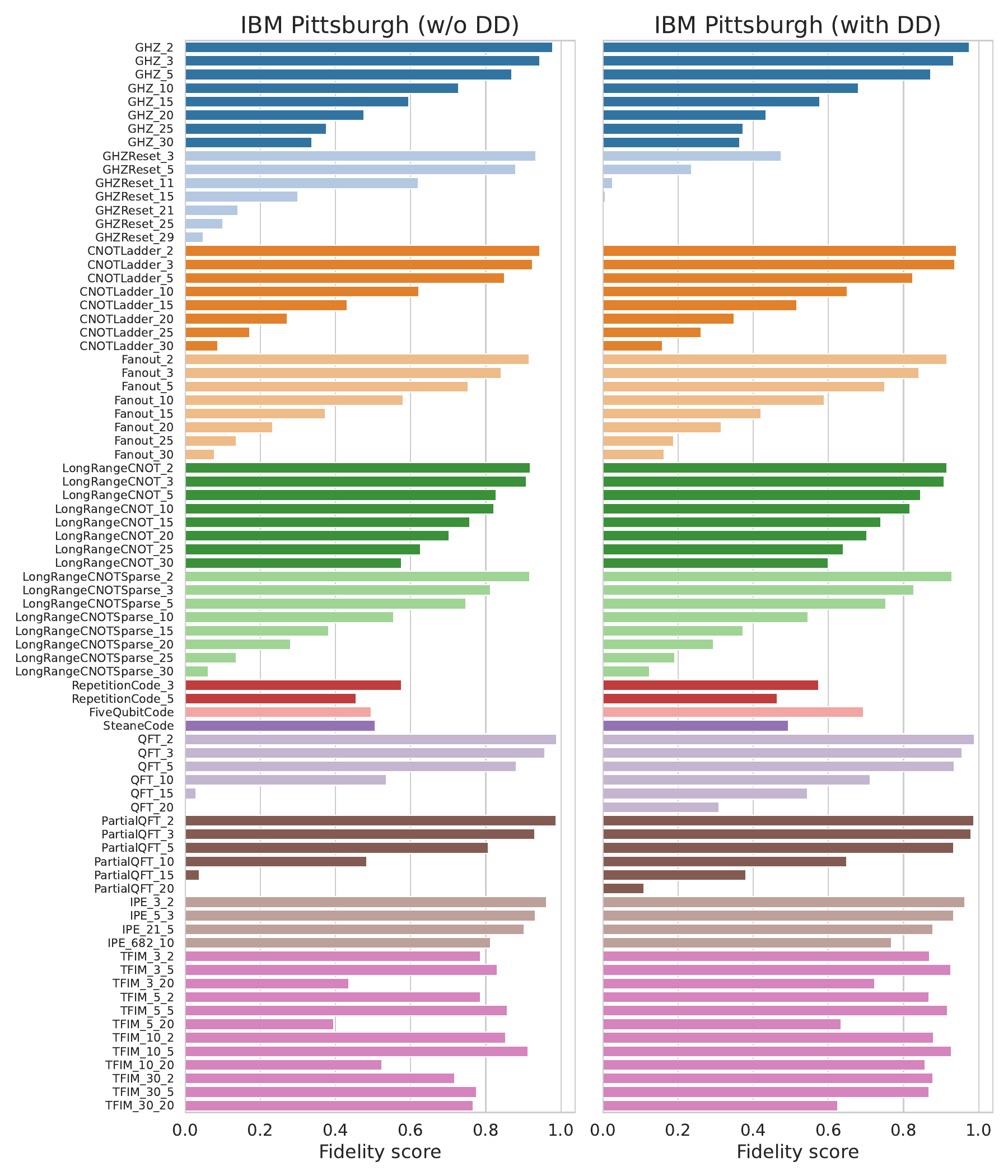}
    \caption{Fidelity scores on IBM Pittsburgh using benchmarks in \textit{dynamarq}.
}
\end{figure*}

\begin{figure*}
    \centering
    \includegraphics[width=\textwidth]{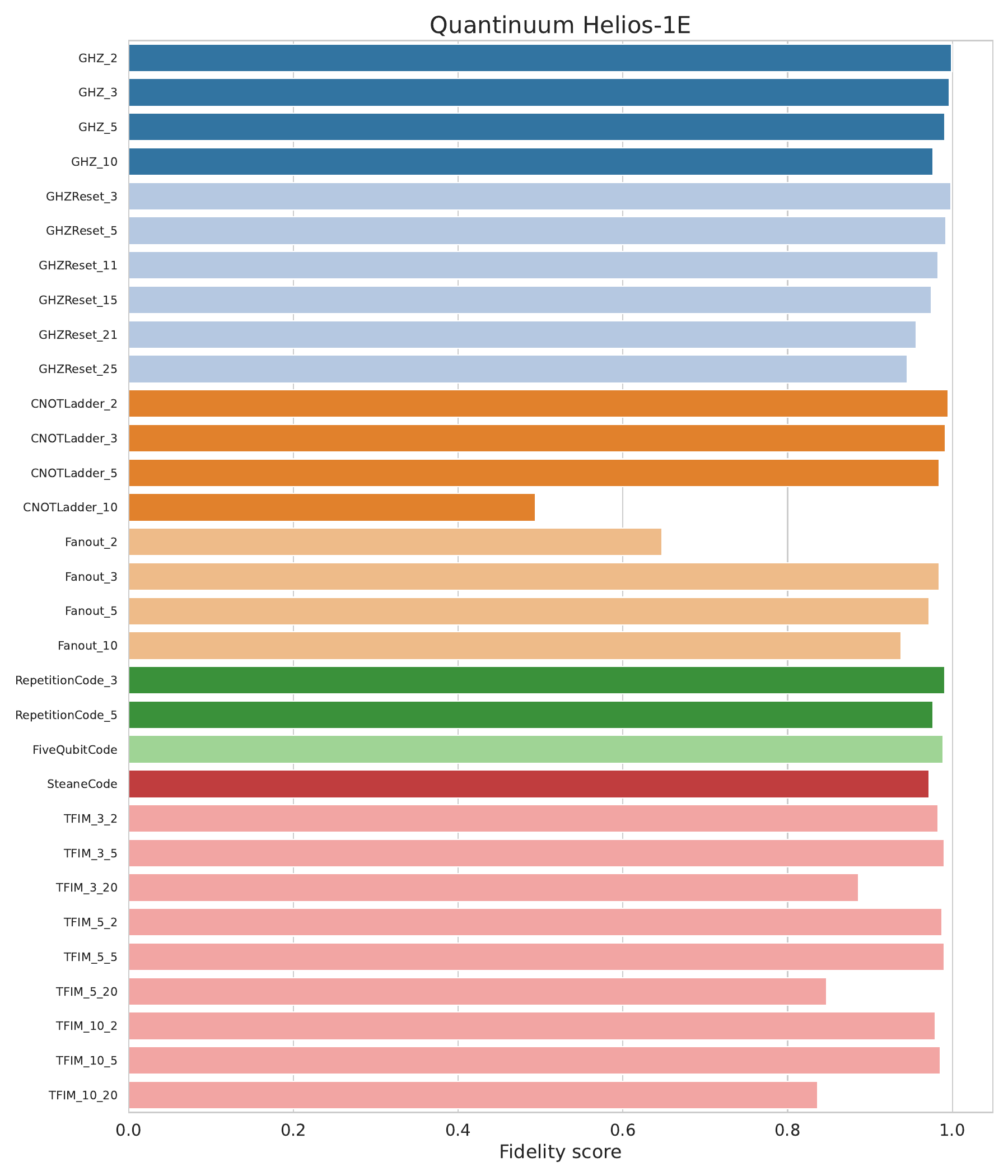}
    \caption{
    Fidelity scores on Quantinuum Helios-1E emulator using benchmarks in \textit{dynamarq}.
    }
\end{figure*}

\bibliography{main}

\end{document}